\documentclass[twocolumn]{aastex631}

\usepackage{amsmath}
\usepackage{float}
\usepackage{threeparttable}
\usepackage{enumitem}
\usepackage{CJKutf8}

\newcolumntype{s}{>{\columncolor{gray!10}} p{3cm}} 

\newcommand{\mergecell}[3]{
  \ifnum#1=1
    \textbf{#3}
  \else
    \ifnum#1=0
      \multicolumn{1}{c|}{\textbf{#3}}
    \fi
  \fi
}

\newcommand{\ContinuedFloat}{}

\newcommand{\lineunder}[1]{\cline{#1}}

\shorttitle{\texttt{ChronoFlow}: A Data-Driven Model for Gyrochronology}
\shortauthors{Van-Lane et al.}

\graphicspath{{./}{images/}}
\begin{document}

\makeatletter
\renewcommand{\@journalinfo}{Accepted for publication in ApJ}
\makeatother

\received{December 5, 2024}
\accepted{March 18, 2025}

\title{\texttt{ChronoFlow}: A Data-Driven Model for Gyrochronology}

\author[0009-0009-4567-9946]{Phil R. Van-Lane}
\affiliation{David A. Dunlap Department of Astronomy \& Astrophysics, University of Toronto, Toronto, ON, Canada}
\affiliation{Dunlap Institute of Astronomy \& Astrophysics, University of Toronto, Toronto, ON, Canada}

\author[0000-0003-2573-9832]{Joshua S. Speagle \begin{CJK*}{UTF8}{gbsn}(沈佳士)\end{CJK*}}
\affiliation{Department of Statistical Sciences, University of Toronto, Toronto, ON M5S 3G3, Canada}
\affiliation{David A. Dunlap Department of Astronomy \& Astrophysics, University of Toronto, Toronto, ON, Canada}
\affiliation{Dunlap Institute of Astronomy \& Astrophysics, University of Toronto, Toronto, ON, Canada}
\affiliation{Data Sciences Institute, University of Toronto, Toronto, ON, Canada}

\author[0000-0003-3734-8177]{Gwendolyn M. Eadie}
\affiliation{Department of Statistical Sciences, University of Toronto, Toronto, ON M5S 3G3, Canada}
\affiliation{David A. Dunlap Department of Astronomy \& Astrophysics, University of Toronto, Toronto, ON, Canada}
\affiliation{Data Sciences Institute, University of Toronto, Toronto, ON, Canada}

\author[0000-0001-7371-2832]{Stephanie T. Douglas}
\affiliation{Department of Physics, Lafayette College, Easton, PA, United States}

\author[0000-0002-1617-8917]{Phillip A. Cargile}
\affiliation{Center for Astrophysics | Harvard \& Smithsonian, Cambridge, MA 02138, USA}

\author[0000-0002-2250-730X]{Catherine Zucker}
\affiliation{Center for Astrophysics | Harvard \& Smithsonian, Cambridge, MA 02138, USA}

\author[0000-0003-4769-3273]{Yuxi (Lucy) Lu}
\affiliation{Department of Astronomy, The Ohio State University, Columbus, 140 W 18th Ave, OH 43210, USA}
\affiliation{Center for Cosmology and Astroparticle Physics (CCAPP), The Ohio State University, 191 W. Woodruff Ave., Columbus, OH 43210, USA}

\author[0000-0003-4540-5661]{Ruth Angus}
\affiliation{American Museum of Natural History, Central Park West, Manhattan, NY, USA}
\affiliation{Center for Computational Astrophysics, Flatiron Institute, Manhattan, NY, USA}

\begin{abstract}

Gyrochronology is a technique for constraining stellar ages using rotation periods, which change over a star's main sequence lifetime due to magnetic braking. This technique shows promise for main sequence FGKM stars, where other methods are imprecise. However, the observed dispersion in rotation rates for similar coeval stars has historically been difficult to characterize. To properly understand this complexity, we have assembled the largest standardized data catalog of rotators in open clusters to date, consisting of $\approx$ 7,600 stars across 30 open clusters/associations spanning ages of 1.5 Myr to 4 Gyr. We have also developed \texttt{ChronoFlow}: a flexible data-driven model which accurately captures observed rotational dispersion. We show that \texttt{ChronoFlow} can be used to accurately forward model rotational evolution, and to infer both cluster and individual stellar ages. We recover cluster ages with a statistical uncertainty of 0.06 dex ($\approx$ 15\%), and individual stellar ages with a statistical uncertainty of 0.7 dex. Additionally, we conducted robust systematic tests to analyze the impact of extinction models, cluster membership, and calibration ages. These contribute an additional ~0.06 dex of uncertainty in cluster age estimates, resulting in a total error budget of 0.08 dex. We apply \texttt{ChronoFlow} to estimate ages for M34, NGC 2516, NGC 6709, and the Theia 456 stellar stream. Our results show that \texttt{ChronoFlow} can precisely estimate the ages of coeval stellar populations, and constrain ages for individual stars. Furthermore, its predictions may be used to inform physical spin down models. \texttt{ChronoFlow} is publicly available at \url{https://github.com/philvanlane/chronoflow}.

\end{abstract}

\section{Introduction} \label{sec: intro}

Stellar ages are critically important to astronomy on a wide range of scales, from exoplanet evolution models to galactic archaeology (e.g., \citealt{Soderblom...AgesOfStars...2010ARA&A..48..581S}). Since its introduction by  \citet{1964ApJ...140..544D}, isochrone fitting has been the most ubiquitous technique for dating stars of a known coeval population. This is done by comparing the observed photometry for stars of various masses against theoretical stellar evolution models, e.g. on a color-magnitude diagram (CMD). However, since stars evolve very little on a CMD over their main sequence (MS) lifetimes, isochrone fitting is imprecise without stars on the giant branch that can be used to locate the MS turn off.

Asteroseismology (e.g., \citealt{2007A&ARv..14..217C, SilvaAguirre..asteroseismology...2017ApJ...835..173S}) has become another popular stellar dating technique, which uses pulsations caused by pressure and gravity waves within stars to infer their internal composition and therefore evolutionary stage. However, pulsations are difficult to detect in low mass MS stars, since the signal magnitude is directly related to the size of the star.

Lithium depletion, which occurs via nuclear reactions in stellar cores, offers yet another way to estimate ages. Lower-mass stars deplete lithium slower because they are cooler, so the Lithium Depletion Boundary (LDB; i.e. the lowest mass stars that have fully consumed their lithium) serves as an age indicator for stellar populations. Analyzing the LDB (e.g., \citealt{2014EAS....65..289J, 2017AA...602A..63B, 2019AJ....158..163D,Jeffries...LiDepAges...2023MNRAS.523..802J}) has shown promise for stars and clusters $\lesssim$ 1 Gyr, but requires spectroscopy to measure Lithium equivalent widths ($EW_{Li}$), and is not well constrained at older ages due to the flattening of Lithium depletion. \cite{Jeffries...LiDepAges...2023MNRAS.523..802J} used an empirical model to estimate individual stellar ages and cluster ages up to 6 Gyr, and achieved precisions of 0.1 dex in the best-constrained cases (M dwarfs $<$ 100 Myr) but struggled to recover training ages at $>$ 1 Gyr.

A stellar dating technique that has proven useful for low mass MS stars is gyrochronology. Gyrochronology is applicable to any stars with a convective envelope ($\sim$ late F and GKM classes, including fully convective M dwarfs), because these stars lose angular momentum over time due to the ejection of material via magnetized winds. This angular momentum loss is observable as a decrease in rotation rate. This theory of ``magnetic braking'' was first proposed by \citet{Schatzman...1962AnAp...25...18S}, and refined further by e.g., \cite{Kraft...spindown...1967ApJ...150..551K}, \cite{Wilson...spindown...1966ApJ...144..695W}, and \cite{WeberDavis...magneticbraking...1967ApJ...148..217W}. \cite{Skumanich...1972ApJ...171..565S} proposed a quantitative empirical model describing this stellar spin down: $P_{rot} \propto \tau^{\frac{1}{2}}$ (where $P_{rot}$ is the rotation period and $\tau$ is the age of a star).

Gyrochronology took another step forward with foundational papers by \cite{2003ApJ...586..464B, 2007ApJ...669.1167B}, who proposed a new empirical model featuring ``fast'' and ``slow'' evolutionary tracks. In these models, the stellar spin down (change in $P_{rot}$) depends on the color (B-V) and age of a star, as well as scaling factors. \citet{2010ApJ...722..222B} and \citet{BarnesKim...2010ApJ...721..675B} developed this further into spindown models dependent on the Rossby number $R_o$ (the ratio of rotation period to convective turnover time) of a star.

Recently, space-based missions such as \textit{Kepler} (\citealt{Kepler...2010Sci...327..977B}), \textit{K2} (\citealt{K2...Howell...2014PASP..126..398H}) and the \textit{Transiting Exoplanet Survey Satellite (TESS; \citealt{TESS...2015JATIS...1a4003R})} have yielded abundant time series photometry and subsequently rotation period measurements. This has enabled a wealth of gyrochronology research, from the enhancement of theoretical spindown models (e.g., \citealt{VanSaders...weakenedbraking...2016Natur.529..181V,SpadaLazafame...spindown...2020A&A...636A..76S,Gossage...brakingmodels...2021ApJ...912...65G}) to the development of data-driven empirical models (e.g., \citealt{Angus...asteroseismology...2015MNRAS.450.1787A,Angus...gyrowithiso...2019AJ....158..173A,Curtis...2019...rotcatalogue,BoyleBouma...gyro.Aper...2022arXiv221109822B}).

One important consideration is the age calibration required to build empirical models. Historically, isochrone fitting to open clusters has been the predominant method (e.g., \citealt{GodoyRivera...2021...rotcatalogue,Long_gyrocatalogue...2023ApJS..268...30L}), which has become increasingly reliable with the precise photometry produced by \textit{Gaia} DR2 \citep{GaiaDR2...documentation...2018gdr2.reptE....V} and beyond. Some studies have also applied asteroseismology (e.g., \citealt{Angus...asteroseismology...2015MNRAS.450.1787A,VanSaders...weakenedbraking...2016Natur.529..181V,Hall...WeakenedBraking...2021NatAs...5..707H}) and wide binaries (e.g., \citealt{Pass...spindown...2022ApJ...936..109P,SilvaBeyer...OldGyroBreakdown...2022arXiv221001137S,Chiti...ConvectiveBounary...2024arXiv240312129C}) as age calibration mechanisms.

Across all of these physical and empirical studies, however, many nuances of the stellar rotational evolution evade simple analytical models. These include:

\begin{enumerate}
    \item Weakened magnetic braking (predominantly in K dwarfs), and a change in rotational evolution of field age stars \citep{Angus...asteroseismology...2015MNRAS.450.1787A,VanSaders...weakenedbraking...2016Natur.529..181V,Curtis...2019...rotcatalogue,Hall...WeakenedBraking...2021NatAs...5..707H,Metcalfe...WeakenedBraking...2022ApJ...933L..17M,SilvaBeyer...OldGyroBreakdown...2022arXiv221001137S}.
    \item Distinct differences in spindown behaviour between partially convective and fully convective M dwarfs \citep{Chiti...ConvectiveBounary...2024arXiv240312129C,Lu...ConvectiveBoundary...2024NatAs...8..223L}.
    \item Separation of fully convective M dwarfs into fast and slow rotators \citep{Garraffo...BimodalSpindown...2018ApJ...862...90G,Pass...spindown...2022ApJ...936..109P}.
    \item Observational scatter in $P_{rot}$ for stars of similar ages and masses/temperatures/colors, particularly at young ages, has been difficult to reproduce with physical or empirical models so far (e.g., \citealt{Curtis...2019...rotcatalogue,Curtis2020_gyro,Bouma...gyrointerp...2023ApJ...947L...3B,Lu...Day&Age...2024AJ....167..159L}). 
\end{enumerate}

\noindent These challenges necessitate a completely flexible, data-driven model that can properly capture these observational trends.

To achieve  this flexibility, we previously presented the first machine learning (ML) based gyrochronology model proof of concept \citep{Van-Lane...icml2023}, and two other recent studies have also probed the flexibility of empirical models. \cite{Bouma...gyrointerp...2023ApJ...947L...3B} developed an interpolation framework which fits the slow rotator sequence to a 7th order polynomial with Gaussian scatter, and the fast rotator sequence as a ``residual'' (compared to the slow sequence mean model) combination of normal and uniform distributions. \cite{Lu...Day&Age...2024AJ....167..159L} implemented a Gaussian Process gyrochronology model, calibrated against both open clusters and gyro-kinematic field star ages (which are derived using an age-velocity-dispersion relation for groups of stars with similar absolute magnitudes, rotation periods, Rossby numbers and effective temperatures). While these methods capture some of the dispersion seen in observations, they still require constraints on the models' functional form.

\begin{figure}
    \centering
    \includegraphics[width=0.48\textwidth]{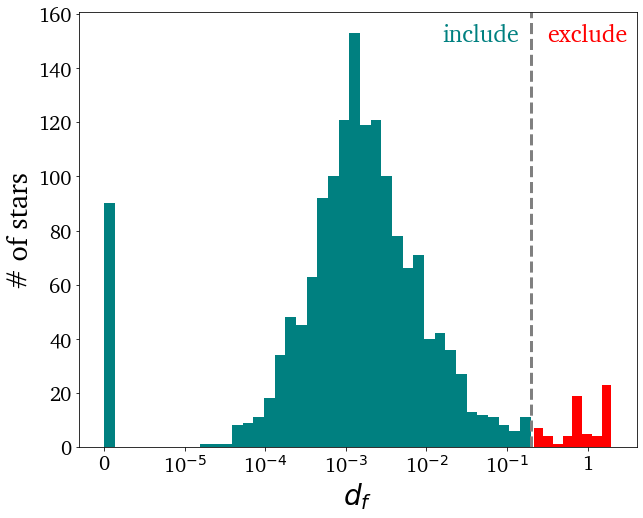}
    \caption{Distribution of the fractional difference ($d_f$) in $P_{rot}$ for stars having two measurements. The value we use as a quality cut (0.2) is illustrated by the dashed grey line. The value of 0.2 was chosen to exclude the tail of stars with very discrepant $P_{rot}$ measurements, while keeping the large majority of these stars.}
    \label{fig:prot_2_df_hist}
\end{figure}

In this work we present our updated model: \texttt{ChronoFlow}, which \textit{models rotational evolution flexibly} with a neural network, and accurately captures observational scatter. We have expanded on our data catalog from \cite{Van-Lane...icml2023} by: (i) including literature rotation periods for 30 open clusters and associations; (ii) standardizing to Gaia DR3 photometry; and (iii) de-reddening using 3D dustmaps from \cite{Edenhofer2023...dustmap...2023arXiv230801295E} and \cite{Bayestar19...dustmap...2019ApJ...887...93G}. We have also optimized our training architecture and enhanced the model itself, which now incorporates photometric uncertainties and cluster membership probabilities.

The rest of the paper is structured as follows. \S\ref{sec: data} outlines how we have assembled our final catalog, crossmatched all rotation periods to Gaia DR3 \citep{GaiaDR3...2023A&A...674A...1G}, applied photometric uncertainties, and assigned cluster membership probabilities to all stars. \S\ref{sec: methods} describes the statistical framework for our model, and \S\ref{sec:chronoflow} details the model itself. \S\ref{sec: results} presents our results, including age recovery tests. In \S\ref{sec: discussion} we describe all of the systematic uncertainties that we have characterized, and quantify our model's performance. We outline applications for \texttt{ChronoFlow} in \S\ref{sec:applications}, \textbf{including age estimates} for five clusters and one stellar stream, and we conclude in \S\ref{sec: conclusions}.

\section{Catalog of Open Cluster Rotators} \label{sec: data}

We conducted a literature search for stellar rotation periods in 30 open clusters and associations, and our compiled catalog is presented in Table \ref{tab:prot_source_table}. It includes four key types of data; we outline these briefly here and describe them further in \S\ref{subsec:prot}-\ref{subsec: cluster_props}.

\begin{enumerate}
    \item \textit{Rotation periods}. We compiled rotation periods ($P_{rot}$) from 16 literature catalogs. Many of these are themselves compilations, so we trace all rotation periods back to their original measurements. Details are discussed in \S\ref{subsec:prot}.
    \item \textit{Photometry}. Where Gaia DR3 IDs were not provided in the source catalog, we crossmatched the stars to DR3 using right ascension (R.A.) and declination (Decl.), and compared magnitudes from the source catalog to DR3 where the crossmatch based on R.A. and Decl. was ambiguous. We de-reddened the DR3 photometry using both the \cite{Edenhofer2023...dustmap...2023arXiv230801295E} and \citet{Bayestar19...dustmap...2019ApJ...887...93G} dustmaps, and propagated the parallax and extinction errors to our de-reddened colors. These details are discussed in \S\ref{subsec: photometry}.
    \item \textit{Cluster membership probability}. We assigned a quantitative cluster membership probability to each star. Our primary source for this was HDBScan \citep{HDBSCAN...McInnes2017}, with which the probabilities were computed as described in \S2.1 of \citet{Douglas...SouthClusters...2024ApJ...962...16D}. We reverted to the membership probabilities from the source catalogs where HDBScan probabilities were not available. Details are discussed in \S\ref{subsec:pclmem}. 
    \item \textit{Cluster ages}. We conducted a literature review of age estimates for the clusters in our catalog; these are compiled in Table \ref{tab:all_lit_ages}.
\end{enumerate}

\subsection{Rotation periods} \label{subsec:prot}

We started with 16,340 rotation periods compiled from all sources in Table \ref{tab:prot_source_table}. After trimming this list based on quality cuts recommended in the source papers and single-star criteria from each data catalog, 12,039 rotation periods remained. The details of these cuts are described in Appendix \ref{app:vetting_all}. Since many of our clusters are in multiple data catalogs, there were two types of measurement duplication in these 12,039 periods:
\begin{enumerate}
    \item Rotation period measurements cited in multiple data catalogs.
    \item Multiple rotation period measurements for the same DR3 source.
\end{enumerate}
To handle the first case, we identified the source of rotation measurement for all stars across the catalogs. In 666 cases, the same $P_{rot}$ measurement was cited in multiple catalogs (see Appendix \ref{sec:app_prot_notes}), and 11,310 out of our 12,039 rotation periods were \textit{unique} measurements.

For the second case, we used the results of our Gaia DR3 crossmatch (see \S \ref{subsec: photometry}) to identify stars with rotation periods measured in multiple studies. We found that of the 11,310 measurements, there were 8,490 unique DR3 sources. Of these, 6,348 had a single $P_{rot}$ measurement, 1,685 stars had two measurements, and 457 stars had three measurements. We include all of the stars with a single $P_{rot}$ in our catalog. For the stars with two $P_{rot}$ measurements, we drop those having a fractional difference ($d_f$) of $\geq$ 20\% from our catalog (77 stars). Figure \ref{fig:prot_2_df_hist} visualizes this cut. While there was evidence that some of these might have been harmonics (where one $P_{rot}$ is twice or half the value of the other), most did not appear to be. For stars having a fractional difference $<$ 20\%, we took the average of the two as our $P_{rot}$.

For stars with three $P_{rot}$ measurements, we first computed the fractional standard deviation of the three values. If it was $\leq$ 2\%, we used the average. If not, we compared the difference between the maximum value and middle value ($d_{max}$) to the difference between the minimum value and the middle value ($d_{min}$). If $d_{max} > 3d_{min}$ or vice versa, we discarded the outlier. We then considered the fractional difference between the two remaining values. If this was $\geq$ 20\%, we exclude the star from our catalog. If it was $<$ 20\%, we used the average of the two as our $P_{rot}$. If $d_{max}$ and $d_{min}$ did not differ by more than $3\times$ in either direction, we excluded the star if the fractional standard deviation of the three values was $\geq$ 20\%, and used the average $P_{rot}$ otherwise.

\onecolumngrid

\begin{longtable}{|l|l|p{3.1cm}|p{4.1cm}|r|}
\caption{Summary of $P_{rot}$ sources. The \textit{Data Catalog} is the literature source, the \textit{$P_{rot}$ Source} is the study in which the $P_{rot}$ was measured, and the \textit{Source Survey} is where the raw data came from. All individual $P_{rot}$ records are provided in the machine-readable version of this table.} \label{tab:prot_source_table} \\

\hline
\textbf{Cluster} & \textbf{Data Catalog} & \textbf{$P_{rot}$ Source} & \textbf{Source Survey} & \textbf{\# of Rotators} \\
\hline
\endfirsthead

\hline
\textbf{Cluster} & \textbf{Data Catalog} & \textbf{$P_{rot}$ Source} & \textbf{Source Survey} & \textbf{\# of Rotators} \\
\hline
\endhead

\hline
\endfoot

\hline
\endlastfoot

$\alpha$ Persei & \cite{BoyleBouma...gyro.Aper...2022arXiv221109822B} & source catalog & TESS & 238 \\
\hline
Collinder 135 & \cite{Douglas...SouthClusters...2024ApJ...962...16D} & source catalog & TESS & 204 \\
\hline
H Persei & \cite{Moraux...gyro...HPer...2013AA...560A..13M} & source catalog & CFHT, Maidanak, ZTSh, Byurakan & 385 \\
\hline
Hyades & \cite{Douglas...Hyades/Praesepe...2019ApJ...879..100D} & source catalog & K2 & 54 \\ 
 & & \cite{Delorme...Hyades...Praesepe...2011MNRAS.413.2218D} & SuperWASP & 18 \\
 & & \cite{Douglas...Hyades...2016ApJ...822...47D} & K2 & 23 \\
 & & \cite{Hartman...HATNet...2011AJ....141..166H} & HATNet & 2 \\
 & & Kundert/Cargile (ASAS)\footnote{sdad} & ASAS & 14 \\
 & & \cite{Radick...Hyades...1987ApJ...321..459R,Radick...Hyades...1995ApJ...452..332R} & Lowell 0.5m & 4 \\
 \lineunder{2-5}
 & \cite{Long_gyrocatalogue...2023ApJS..268...30L} & source catalog & K2 & 88 \\
 & & \cite{Delorme...Hyades...Praesepe...2011MNRAS.413.2218D} & SuperWASP & 9 \\
 & & \cite{Hartman...HATNet...2011AJ....141..166H} & HATNet & 1 \\

\hline
IC 2391 & \cite{Douglas...SouthClusters...2024ApJ...962...16D} & source catalog & TESS & 96 \\
\hline
IC 2602 & \cite{Douglas...SouthClusters...2024ApJ...962...16D} & source catalog & TESS & 175 \\
\hline
M34 & \cite{Meibom...M34...2011ApJ...733..115M} & source catalog & WIYN & 68 \\
\hline
M35 & \cite{Long_gyrocatalogue...2023ApJS..268...30L} & \cite{Libralato...M35...2016MNRAS.456.1137L} & Kepler/K2 & 265 \\
 &  & \cite{Meibom...M35...2009ApJ...695..679M} & WIYN & 126 \\
 &  & \cite{Nardiello...M35...NGC2158...2015MNRAS.447.3536N} & Schmidt Telescope & 22 \\
 &  & \cite{SoaresFurtado...M35...NGC2158...2020ApJS..246...15S} & K2 & 124 \\
 \hline
M37 & \cite{GodoyRivera...2021...rotcatalogue} & \cite{JH+2009b} & K2 & 340 \\
\hline
M50 & \cite{GodoyRivera...2021...rotcatalogue} & \cite{Irwin...M50...2009MNRAS.392.1456I} & Blanco telescope at CTIO & 577 \\
\hline
M67 & \cite{Barnes...M67...2016ApJ...823...16B} & source catalog & K2 & 17 \\
\lineunder{2-5}
 & \cite{Dungee...gyro.M67...2022ApJ...938..118D} & source catalog & CFHT & 215 \\
 \lineunder{2-5}
 & \cite{Long_gyrocatalogue...2023ApJS..268...30L} & source catalog & K2 & 62 \\
 & & \cite{Dungee...gyro.M67...2022ApJ...938..118D} & CFHT & 45 \\
\hline
NGC 1647 & \cite{Long_gyrocatalogue...2023ApJS..268...30L} & source catalog & K2 & 35 \\
\hline
NGC 1750 & \cite{Long_gyrocatalogue...2023ApJS..268...30L} & source catalog & K2 & 39 \\
\hline
NGC 1758 & \cite{Long_gyrocatalogue...2023ApJS..268...30L} & source catalog & K2 & 21 \\
\hline
NGC 1817 & \cite{Long_gyrocatalogue...2023ApJS..268...30L} & source catalog & K2 & 60 \\
\hline
NGC 2281 & \cite{Fritzewski...gyro...NGC2281...2023AA...674A..152F} & source catalog & AIP STELLA I, IAC & 116 \\
\hline
NGC 2451A & \cite{Douglas...SouthClusters...2024ApJ...962...16D} & source catalog & TESS & 129 \\
\hline
NGC 2516 & \cite{GodoyRivera...2021...rotcatalogue} & \cite{Irwin...NGC2516...2007MNRAS.377..741I} & Blanco telescope at CTIO & 327 \\
\lineunder{2-5}
 & \cite{Fritzewski...NGC2516...2020AA...641A..51F} & \cite{Irwin...NGC2547...2008MNRAS.383.1588I} & Blanco telescope at CTIO & 309 \\
 & & source catalog & Yale telescope at CTIO & 303 \\
 \hline
NGC 2548 & \cite{Barnes...gyro...NGC2548...2015AA...583A..73B} & source catalog & AIP STELLA I, IAC & 59 \\
\hline
NGC 2547 & \cite{Douglas...SouthClusters...2024ApJ...962...16D} & source catalog & TESS & 80 \\
\lineunder{2-5}
 & \cite{GodoyRivera...2021...rotcatalogue} & \cite{Irwin...NGC2547...2008MNRAS.383.1588I} & MPG/ESO & 109 \\
 \hline
NGC 2548 & \cite{Barnes...gyro...NGC2548...2015AA...583A..73B} & source catalog & AIP STELLA I, IAC & 59 \\
\hline
NGC 3532 & \cite{Fritzewski...gyro.NGC3532...2021AA...656A.103F} & source catalog & Yale telescope at CTIO & 228 \\
\hline
NGC 6811 & \cite{Curtis2020_gyro} & \cite{Curtis...2019...rotcatalogue} & Kepler & 163 \\
\lineunder{2-5}
 & \cite{GodoyRivera...2021...rotcatalogue} & \cite{Curtis...2019...rotcatalogue} & Kepler & 32 \\
 & & unknown & & 3 \\
 & & \cite{AS+19_Kepler_catalog} & Kepler & 60 \\
 & & \cite{AS+21_Kepler_catalog} & Kepler & 118 \\
\lineunder{2-5}
 & \cite{Long_gyrocatalogue...2023ApJS..268...30L} & source catalog & Kepler & 220 \\
\hline
NGC 6819 & \cite{Curtis2020_gyro} & \cite{SM+15_NGC6819} & Kepler & 28 \\
\lineunder{2-5}
 & \cite{Long_gyrocatalogue...2023ApJS..268...30L} & source catalog & Kepler & 151 \\
\hline
NGC 6866 & \cite{Long_gyrocatalogue...2023ApJS..268...30L} & source catalog & Kepler & 124 \\
\hline
NGC 752 & \cite{Curtis2020_gyro} & \cite{Agueros...NGC752...2018ApJ...862...33A} & PTF & 8 \\
\hline
Pleiades & \cite{Curtis2020_gyro} & \cite{Rebull...Pleiades...2016AJ....152..113R} & K2 & 304 \\
\lineunder{2-5}
& \cite{GodoyRivera...2021...rotcatalogue} & \cite{Rebull...Pleiades...2016AJ....152..113R} & K2 & 597 \\
\lineunder{2-5}
 & \cite{Long_gyrocatalogue...2023ApJS..268...30L} & source catalog & K2 & 552 \\
 & & \cite{Hartman...Pleiades...2010MNRAS.408..475H} & HAT-9/HAT-10 & 104 \\
\hline
Praesepe & \cite{GodoyRivera...2021...rotcatalogue} & \cite{Rebull...Praesepe...2017} & K2 & 588 \\
\lineunder{2-5}
 & \cite{Long_gyrocatalogue...2023ApJS..268...30L} & source catalog & K2 & 692 \\
 & & \cite{Delorme...Hyades...Praesepe...2011MNRAS.413.2218D} & SuperWASP & 30 \\
\lineunder{2-5}
 & \cite{Rampalli...Praesepe...2021ApJ...921..167R} & source catalog & K2 & 512 \\
\hline
$\rho$ Ophiucus & \cite{Long_gyrocatalogue...2023ApJS..268...30L} & source catalog & K2 & 208 \\
\lineunder{2-5}
 & \cite{Rebull...gyro.USco.ROph...2018AJ....155..196R} & source catalog & K2 & 72 \\
\hline
Ruprecht 147 & \cite{Curtis2020_gyro} & source catalog & K2;PTF & 27 \\
\lineunder{2-5}
 & \cite{Long_gyrocatalogue...2023ApJS..268...30L} & source catalog & K2 & 34 \\
 & & \cite{Curtis2020_gyro} & K2;PTF & 7 \\
\hline
Taurus & \cite{Long_gyrocatalogue...2023ApJS..268...30L} & source catalog & K2 & 92 \\
\hline
Upper Scorpius & \cite{Long_gyrocatalogue...2023ApJS..268...30L} & source catalog & K2 & 583 \\
\lineunder{2-5}
 & \cite{Rebull...gyro.USco.ROph...2018AJ....155..196R} & source catalog & K2 & 625 \\
\hline\hline
\textbf{Total} & \textbf{16 Data Catalogs} & \textbf{37 $P_{rot}$ Sources} & \textbf{20 Source Surveys} & \textbf{12,039 Periods} \\
\hline

\end{longtable}

\vspace{-8pt}

\begin{tablenotes} 
\item[1] $^1$ These data were not published; see footnote 5 in \S2.1 of \cite{Douglas...Hyades/Praesepe...2019ApJ...879..100D}.
\end{tablenotes}


\vspace{10pt}

\twocolumngrid

With this process, only one star with three measured rotation periods was excluded from our catalog, and we were left with 8,412 stars with reliable rotation periods in total. We describe edge cases that we encountered when compiling these rotation periods in Appendix \ref{sec:app_prot_notes}.

\begin{figure*}
    \centering
    \includegraphics[width=0.95\textwidth]{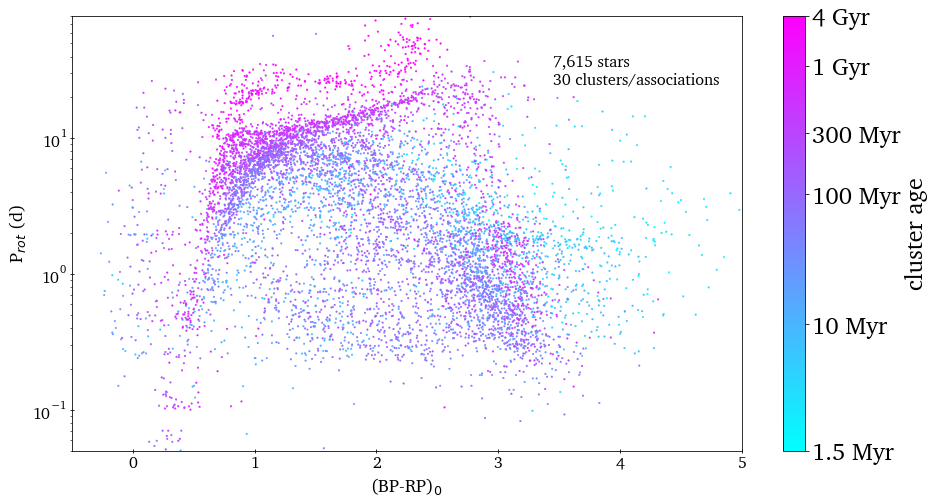}
    \caption{De-reddened color-rotation space for all stars in our final catalog, including the young subgroups of Taurus and Sco-Cen. Stars are color-coded according to the age of their cluster. Older stars appear to have generally converged onto slowly rotating sequences more than younger stars.}
    \label{fig:Prot_all}
\end{figure*}

\subsection{Photometry} \label{subsec: photometry}

For consistency, we use Gaia DR3 BP and RP bands \citep{GaiaBPRP...2021A&A...649A...3R} for all photometry. Of the 12,039 compiled rotation periods, we could not find a DR3 crossmatch for 49 of them. Additionally, 170 of the stars that we did crossmatch belong in Taurus, Upper Scorpius, or $\rho$ Ophiucus, but were not part of the \cite{Ratzenbock...ScoCen_groups...2023AA...677A..59R} or \cite{Krolikowski...Taurus...2021AJ....162..110K} catalogs (which we use as our source of truth for those regions). We successfully crossmatched the other 11,820 rotators, using the steps described in Appendix \ref{app:sec_dr3_crossmatch}, corresponding to 8,412 unique DR3 sources. In our final training catalog we have excluded 21 sources that do not have both BP and RP photometry, three sources from NGC 3532 for which we were unable to calculate de-reddened colors, and an additional 736 having Gaia \texttt{RUWE} values $\geq$ 1.4 (although these are included in the online table). This last criteria is intended to exclude likely binaries that made it through the source catalog cuts, although we recognize that there are caveats. This cut may not capture short-period binaries (e.g., \citealt{ElBadry...GaiaBinaries...2024NewAR..9801694E,Grondin...WDMSCandidates...2024ApJ...976..102G}), and \citet{Penoyre...GaiaBinaries...2022MNRAS.513.5270P} indicate that a cut of 1.25 is more appropriate for DR3. We include an outlier component in our model to account for binaries that remain in our sample, but acknowledge that handling this more thoughtfully in the future could help our model performance (see \S\ref{subsec:dis_binarity}).

Finally, we apply a cut to exclude post-MS stars. While many of the source catalogs have already been vetted for these, we have performed a final CMD check, and excluded an additional 37 stars (see Appendix \ref{app:cluster_summary_plots}). After these cuts, our final catalog consists of 7,615 unique Gaia DR3 MS sources with reliable rotation periods. These are plotted in Figure \ref{fig:Prot_all}. We do not include a cut to remove pre-MS (PMS) stars. Although the rotational behavior of such stars is different than MS stars ``spinning down'', we keep PMS stars in our catalog since they can persist in clusters up to hundreds of Myr in age, and therefore are an important factor in the observed rotational distribution of young clusters.

\begin{figure}
    \centering
    \includegraphics[width=0.45\textwidth]{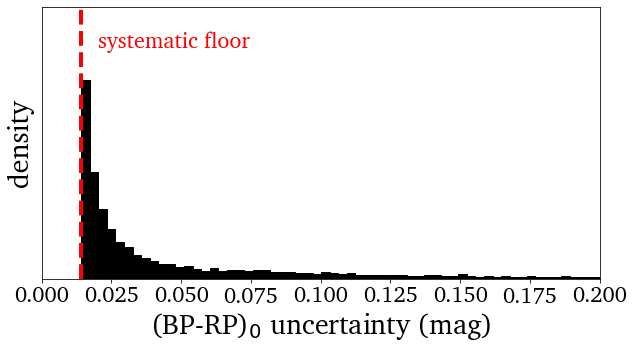}
    \caption{Distribution of $(BP-RP)_0$ uncertainties, including propagation of parallax, dustmap, and conversion errors. There is a minimum value of 0.01415 indicated by the red dashed line; this is the DR3 systematic floor corresponding to errors of 0.01 in the BP and RP bands. This plot is truncated at 0.2 mag, but there is a thin tail that extends to 1.24 mag.}
    \label{fig:bprp0_err_hist}
\end{figure}

\subsubsection{Stellar de-reddening} \label{subsubsec: dered}

The effect of foreground dust needs to be accounted for. Since dust absorbs and scatters more light at short wavelengths (e.g.,  \citealt{Cardelli...extinction...1989ApJ...345..245C}), this \textit{extinction} affects the BP band more than the RP band, and therefore influences $(BP-RP)$. We correct for this by de-reddening every star based on 3D dust maps.

For every star within 1.25 kpc, we used the \citet{Edenhofer2023...dustmap...2023arXiv230801295E} map because it provides parsec-scale distance resolution. \citet{Edenhofer2023...dustmap...2023arXiv230801295E} released the differential extinction per parsec, $E_{XP}$,  measured in an arbitrary reddening unit based on \citet{Zhang...extinction...2023MNRAS.524.1855Z}, which we integrated to the appropriate distance for each cluster. For all stars beyond 1.25 kpc (H Persei, M 37, NGC 1817, NGC 6819, NGC 6866), we used extinctions from the \citet{Bayestar19...dustmap...2019ApJ...887...93G} 3D dust map ($E_{B19}$), which are measured in an arbitrary reddening unit and can be converted to extinction in the Pan-STARRS and 2MASS passbands using the coefficients in their Table 1. 

To convert the arbitrary reddening from \cite{Edenhofer2023...dustmap...2023arXiv230801295E} and \cite{Bayestar19...dustmap...2019ApJ...887...93G} to $E(B-V)$, we used the following:
\begin{equation} \label{eq: EBV_edenhofer}
    E(B-V) = 0.829 \times E_{XP}
\end{equation}
\begin{equation} \label{eq: EBV_bayestar}
    E(B-V) = 0.88 \times E_{B19}
\end{equation}
Eqn. \eqref{eq: EBV_edenhofer} is derived from the combination of $\frac{E_{SFD}}{E_{XP}} = 0.938$\footnote{Ratio derived from comparing the $E_{XP}$ reddening and the $E_{SFD}$ reddening at high latitudes (Xiangyu Zhang 2024, private communication), where the two estimates are expected to agree.}  and $\frac{E(B-V)}{E_{SFD}} = 0.884$ from \cite{Schlafly...dered...2011ApJ...737..103S}, where $E_{XP}$ is the \cite{Zhang...extinction...2023MNRAS.524.1855Z} extinction adopted in \citet{Edenhofer2023...dustmap...2023arXiv230801295E} and $E_{SFD}$ is the reddening unit based on the 2D dust emission map from \cite{Schlegel_1998}. Eqn. \eqref{eq: EBV_bayestar} is derived from combining $E(g-r) = 0.901 \times E_{B19}$ (Eqn. 30 from \citealt{Bayestar19...dustmap...2019ApJ...887...93G}) with $E(B-V) = 0.98 \times E(g-r)$ from \citet{Schlafly...dered...2011ApJ...737..103S}. Once we obtained $E(B-V)$, we converted that to $A_V$ using a differential reddening value of $R_V = 3.1$, and converted $A_V$ to extinction in the Gaia bands $A_G$, $A_{GBP}$, and $A_{GRP}$ using Table 3.1 from \cite{Wang...ext.conversions...2019ApJ...877..116W}.

We used the Gaia DR3 R.A. and Decl. values to query the dust maps using the \texttt{dustmaps}\footnote{\url{https://github.com/gregreen/dustmaps}} package from \citet{Green...dustmaps...2018JOSS....3..695M}. We sampled 10 distances for each star from a normal distribution in parallax, after applying zeropoint corrections using \texttt{gaiadr3\_zeropoint} (\citealt{Lindegren...zeropoint...2021A&A...649A...4L}). For DR3 sources having a relative parallax error of $>$20\%, we instead sampled from a normal distribution in distance representative of the stars in the same cluster having $<$20\% fractional parallax error. We also sampled extinctions from each dustmap, and propagated these uncertainties in distance and extinction through the conversions in Eqns. \eqref{eq: EBV_edenhofer} and \eqref{eq: EBV_bayestar} to get our de-reddened uncertainties. Figure \ref{fig:bprp0_err_hist} shows the final distribution of these $(BP-RP)_0$ uncertainties, truncated at 0.2 mag for ease of visualization.

Here we note that for stars covered by both the \cite{Edenhofer2023...dustmap...2023arXiv230801295E} and \cite{Bayestar19...dustmap...2019ApJ...887...93G} dustmaps, there are systematic differences in the measured extinctions. The impact of this is investigated in \S\ref{sec: discussion} and Appendix \ref{subsec:sys_dustmap_comparison}. 

\subsection{Cluster membership}
\label{subsec:pclmem}

The HDBScan cluster membership probabilities computed by \cite{Douglas...SouthClusters...2024ApJ...962...16D} are available for 79\% of the total number of catalog stars across 27 of our clusters; the percentage per cluster varies. HDBScan was not used for Hyades, Taurus, or Upper Scorpius/$\rho$  Ophiucus because these are all dispersed in the sky; therefore (i) they would require computationally expensive cone searches, and (ii) the assumption of strong  parallax and proper motion definition breaks down, and these associations would require additional treatment.  For all stars without HDBScan values, we use the membership probabilities from the source catalogs where available, and otherwise assign default values according to how the cluster membership was derived (see Appendix \ref{app:cluster_membership_probs_details}).

HDBScan membership probabilities are heavily dependent on the hyperparameters used, so these introduce a systematic uncertainty in our catalog. However, we find that the overall impact on our model is negligible (see \S\ref{sec: discussion}).

\subsection{Cluster ages} \label{subsec: cluster_props}

\begin{figure*}
    \centering
    \includegraphics[width=0.98\textwidth]{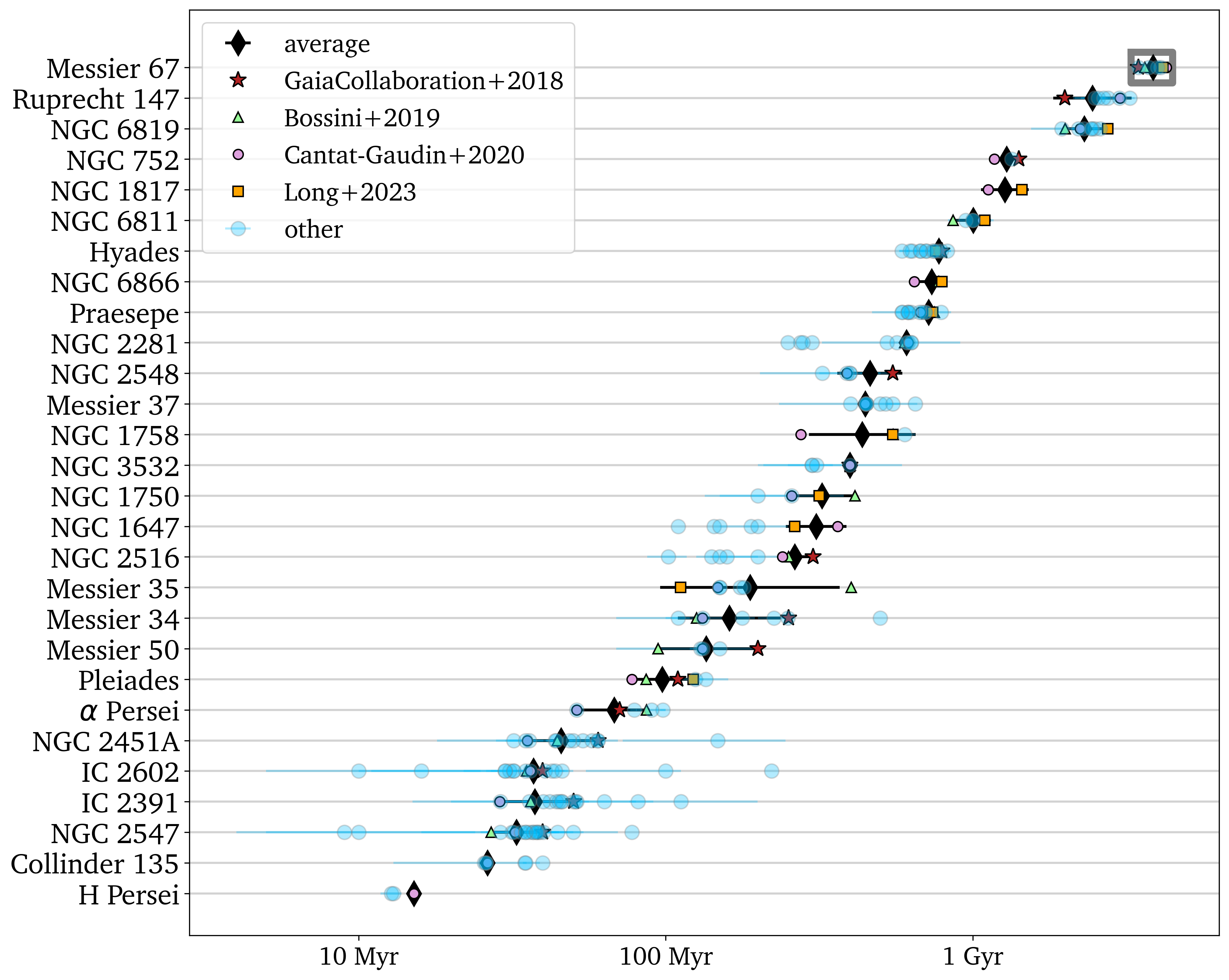}
    \caption{Literature age estimates for our clusters, demonstrating the coverage in our catalog. The values from the four main catalogs are highlighted explicitly as red, green, pink, and orange points, and the average value is shown as a black diamond (with an error bar showing the standard deviation across the four catalogs). Other literature estimates are shown as translucent blue circles (see Figure \ref{fig: M67_lit_ages} as an example). We exclude the young subgroups of Taurus and Upper Scorpius/$\rho$ Ophiucus from this plot.}
    \label{fig: logA_lit_all}
\end{figure*}

\begin{figure*}
    \centering
    \includegraphics[width=0.98\textwidth]{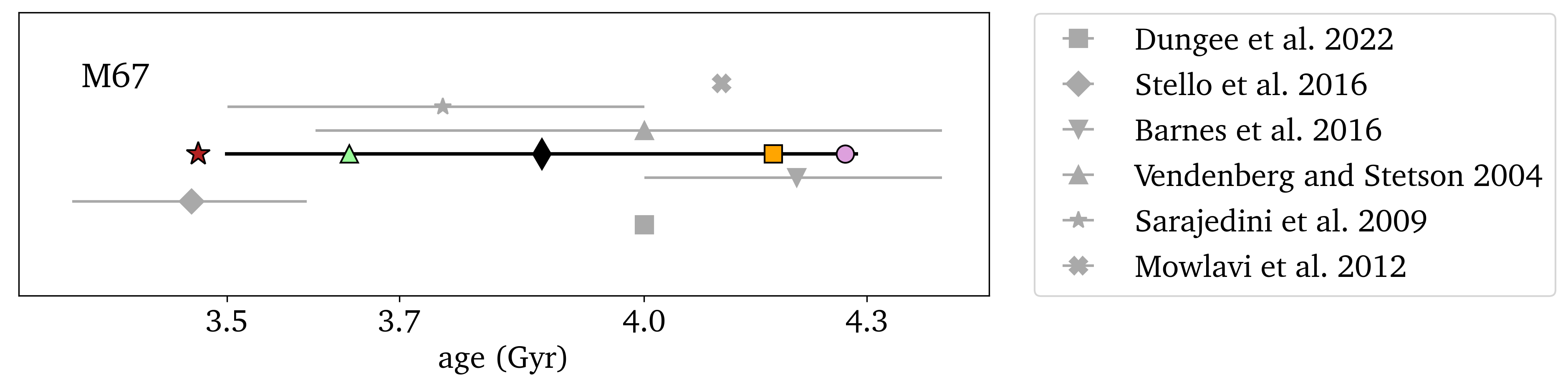}
    \caption{A zoomed in view of M67 from Figure \ref{fig: logA_lit_all}. Age estimates from all sources are shown here in detail, and jittered vertically for ease of visualization.}
    \label{fig: M67_lit_ages}
\end{figure*}

We used four literature catalogs as our primary sources for cluster ages: 
\cite{GaiaCollab...2018AA...616A..10G} (hereafter G+18), 
\cite{Bossini...Catalogue...2019AA...623A.108B} (hereafter B+19),
\cite{CantatGaudin...catalogue...2020AA...640A...1C} (hereafter CG+20), and \cite{Long_gyrocatalogue...2023ApJS..268...30L} (hereafter L+23). Each study implemented PARSEC isochrones to fit for cluster properties. We use the average ages across the four catalogs as the fiducial age for each cluster.

The exceptions to the above are the young associations Taurus and Upper Scorpius, and the young cluster $\rho$ Ophiucus. L+23 compiled rotation periods from all three, and \cite{Rebull...gyro.USco.ROph...2018AJ....155..196R} compile rotation periods from the latter two. Some rotators assigned to Upper Scorpius by L+23 are assigned to $\rho$ Ophiucus by \cite{Rebull...gyro.USco.ROph...2018AJ....155..196R}, and vice versa. However, we ignore these designations and instead use the subgroups and corresponding ages from \cite{Ratzenbock...ScoCen_groups...2023AA...677A..59R}. Similarly, we break the group of Taurus rotators compiled by L+23 into the sub-groups defined by \cite{Krolikowski...Taurus...2021AJ....162..110K}, and use those ages.

We also conducted a literature review of other age estimates for our clusters, which are summarized in Figure \ref{fig: logA_lit_all} and Table~\ref{tab:all_lit_ages}. As an example of the variation among studies, Figure \ref{fig: M67_lit_ages} presents all age estimates for (some ages are jittered vertically for visualization purposes).

\begin{table}[h!]
\centering
\renewcommand{\arraystretch}{1.3}
\begin{tabular}{|l|p{5cm}|}
\hline
\textbf{Field Name} & \textbf{Description} \\ \hline
Cluster & Name of the star cluster. \\ \hline
Source & Paper in which the age estimate was calculated. \\ \hline
Methodology & The technique used to estimate the age of the cluster. \\ \hline
Age & The cluster's age in millions of years. \\ \hline
Age\_err1 & Upper uncertainty on the age in millions of years (when available). \\ \hline
Age\_err2 & Lower uncertainty on the age in millions of years (when available). \\ \hline
\end{tabular}
\caption{This table contains the cluster age estimates summarized in Figure \ref{fig: logA_lit_all}. The full version is available online in machine-readable format.}
\label{tab:all_lit_ages}
\end{table}

\section{Age inference framework} \label{sec: methods}

To use our data catalog in a statistically robust way, we developed a Bayesian framework to describe the inference of stellar ages ($\tau$) given rotation period ($P_{rot}$), de-reddened DR3 color (described as $C_0 = BP_0 - RP_0$), and photometric uncertainty ($\sigma_{C_0}$). In this context, we have implemented Bayes' theorem as:
\begin{equation} \label{eq: bayes_with_var}
    \mathcal{P}(\tau\;|\;P_{rot}, C_0, {\sigma}_{C_0}) = \frac{\mathcal{P}(P_{rot}, {C_0}\;|\;\tau, \sigma_{C_0}) \cdot \mathcal{P}(\tau\;|\;\sigma_{C_0})}{\mathcal{P}(P_{rot},C_0\;|\;\sigma_{C_0})}
\end{equation}

\noindent We have chosen to include $\sigma_{C_0}$ as an independent parameter for ease of computation, but it would also be possible to incorporate it directly within the characterization of $C_0$; both strategies ensure that the impact of color uncertainty is included. We condition all parts of our model on $\sigma_{C_0}$, so do not assume any prior on that parameter.

In \eqref{eq: bayes_with_var}, we can expand $\mathcal{P}(P_{rot}, {C_0}\;|\;\tau, \sigma_{C0})$ as:
\begin{equation}
\begin{aligned} \label{eq: likelihood_expanded}
    \mathcal{P}(P_{rot}, {C_0}\;|\;\tau, \sigma_{C_0}) & = \\
    \mathcal{P}(P_{rot}\;|\;C_0,\sigma_{C_0},&\tau) \cdot \mathcal{P}(C_0\;|\;\sigma_{C_0},\tau)
\end{aligned}
\end{equation}

so that our full equation is:
\begin{equation}
\label{eq: bayes_with_var_expanded}
\begin{aligned}
    \mathcal{P}&(\tau\;|\;P_{rot}, C_0, \sigma_{C_0}) = \\
    &\frac{\mathcal{P}(P_{rot}\;|\; {C_0},\sigma_{C_0},\tau)
    \cdot \mathcal{P}(C_0\;|\;\sigma_{C_0},\tau) \cdot \mathcal{P}(\tau\;|\;\sigma_{C_0})}{\mathcal{P}(P_{rot},C_0\;|\;\sigma_{C_0})}
\end{aligned}
\end{equation}

\noindent We choose to separate the likelihood into two terms in this way so that our model learns the $P_{rot}$ distribution \textit{conditioned} on the other parameters, and not the overall $P_{rot}$ distribution. By doing so, \textit{we mitigate the influence of selection effects in color and age}, and so minimize the impact of observational biases on our age inference.

With our framework in this form, we can evaluate each term on the right independently. $\mathcal{P}(P_{rot},C_0\;|\;\sigma_{C_0})$ in the denominator is the evidence term: this is the integral of the numerator over all possible parameters, and represents the overall likelihood of our observations given our model. Since we are only using a single model, the evidence is a constant scaling term, and the actual value does not matter since we normalize our posteriors to integrate to 1. $\mathcal{P}(\tau\;|\;\sigma_{C0})$ and $\mathcal{P}(C_0\;|\;\sigma_{C_0},\tau)$ are the priors. $\mathcal{P}(\tau\;|\;\sigma_{C0})$ is defined by:
\begin{equation} \label{eq: age_prior}
    \mathcal{P}(\tau\;|\;\sigma_{C0}) \sim \mathcal{U}[1,13800]\;\mathrm{Myr}
\end{equation}

\noindent While this represents the expected distribution of $\tau$ conditioned on $\sigma_{C_0}$, the age of a star does not physically depend on color uncertainty in any way, so we apply a uniform age prior extending from 1 Myr to the age of the universe.

To build the $\mathcal{P}(C_0\;|\;\sigma_{C_0},\tau)$ term, we firstly assume that the observed colour $C_0$ does not depend on the age of a star. While this is not strictly true (the bluest stars turn off the MS relatively quickly and therefore are more likely to be younger), our model is designed to optimize the \textit{conditional} distribution of rotation period, so we ignore this effect and assume:
\begin{equation}
\label{eq:c0_given_age}
    \mathcal{P}(C_0\;|\;\tau) \sim U[-0.05,3.8]
\end{equation}
The limits of -0.05 and 3.8 are chosen based on the de-reddened colors we expect to see in our catalog with 0 photometric error. 

\begin{figure}
    \centering
    \includegraphics[width=0.48\textwidth]{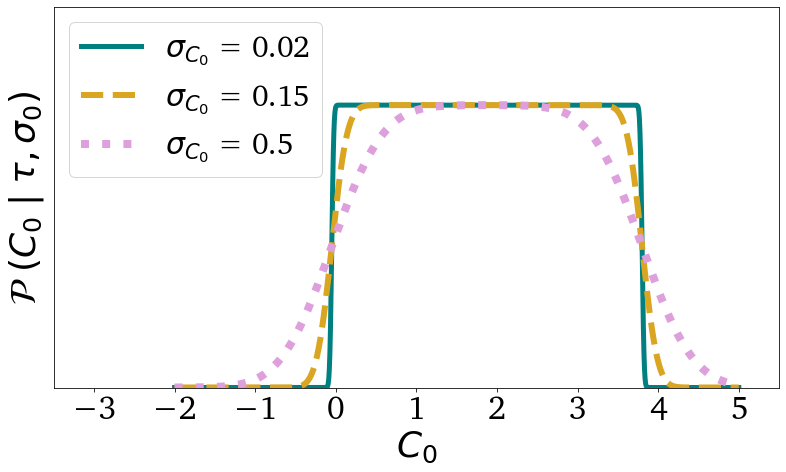}
    \caption{A comparison of the distribution described in Eqn. \eqref{eq:prior_C0_cond_age_err} for different values of $\sigma_{C_0}$. Larger $\sigma_{C_0}$ results in a wider, smoother prior around the expected $C_0$ limits.}
    \label{fig:prior_sigma0_comparison}
\end{figure}

We incorporate $\sigma_{C_0}$ by convolving \eqref{eq:c0_given_age} with a Gaussian of variance $\sigma_{C_0}^2$, so that the functional form becomes:
\begin{equation} \label{eq:prior_C0_cond_age_err}
    \mathcal{P}(C_0\;|\;\tau,\sigma_{C_0}) \sim \mathcal{U}[-0.05,3.8] * \mathcal{N}(0,\sigma_{C_0})
\end{equation}

\noindent which effectively sums the two probability distributions on the right. Figure \ref{fig:prior_sigma0_comparison} illustrates Eqn. \eqref{eq:prior_C0_cond_age_err} as the change in $C_0$ as a function of $\sigma_{C_0}$. As expected, when $\sigma_{C_0}$ is small the uniform $C_0$ distribution dominates, but as $\sigma_{C_0}$ increases then $\mathcal{P}(C_0\;|\;\tau,\sigma_{C_0})$ becomes more Gaussian.

\begin{figure*}
    \centering
    \includegraphics[width=0.9\textwidth]{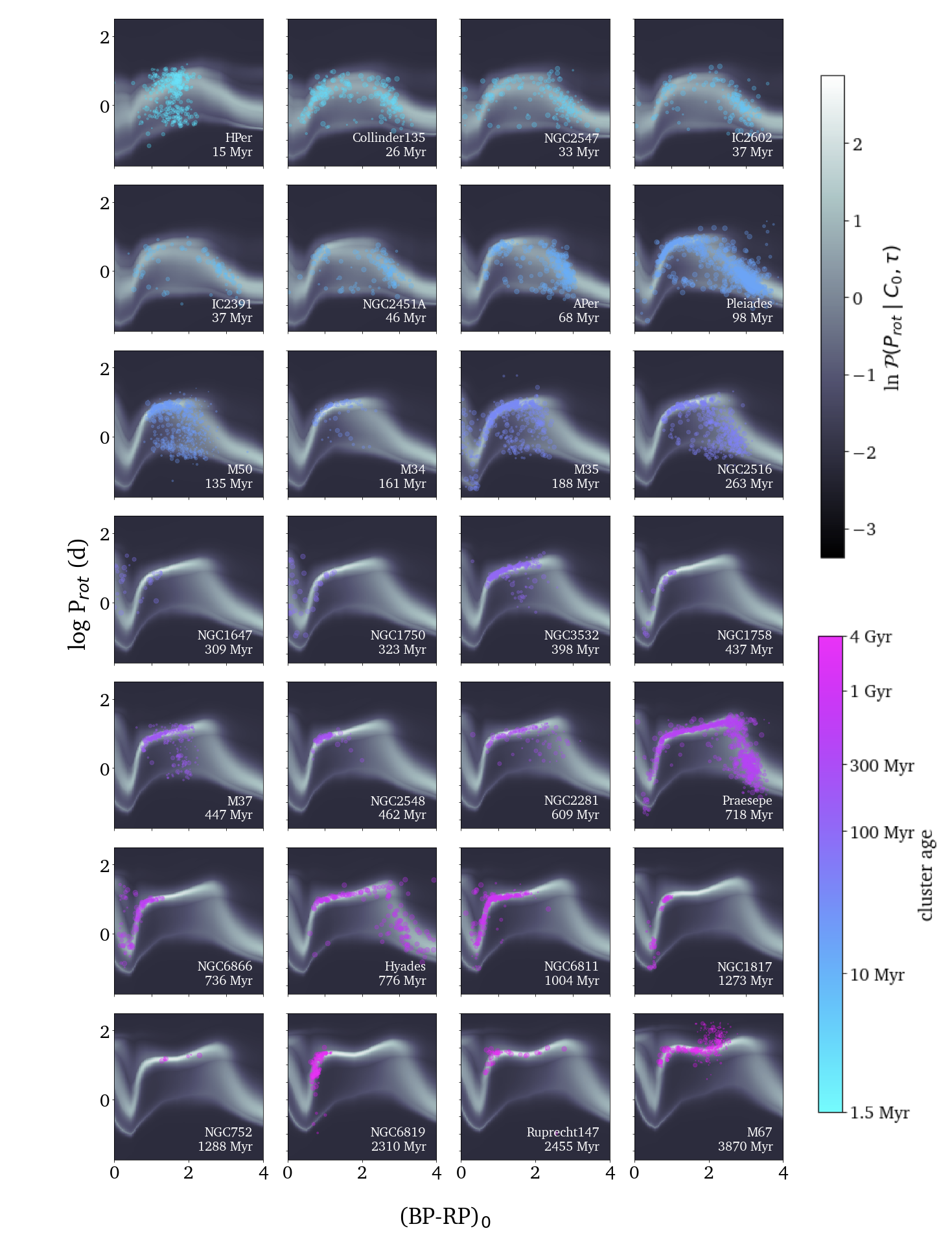}
    \caption{$\mathcal{P}(P_{rot}\;|\;C_0,\sigma_{C_0},\tau)$ (background shading) from \texttt{ChronoFlow} at each cluster's fiducial literature age. Observations from our catalog are plotted as circles (sized by cluster membership probability). We calculate this using representative values for $\sigma_{C_0}$ and $p_{cl}$ of 0.028 (the median across our entire catalog) and 0.9 respectively.}
    \label{fig: nf_dens_old}
\end{figure*}

\begin{figure*}[b]
    \centering
    \includegraphics[width=0.98\textwidth]{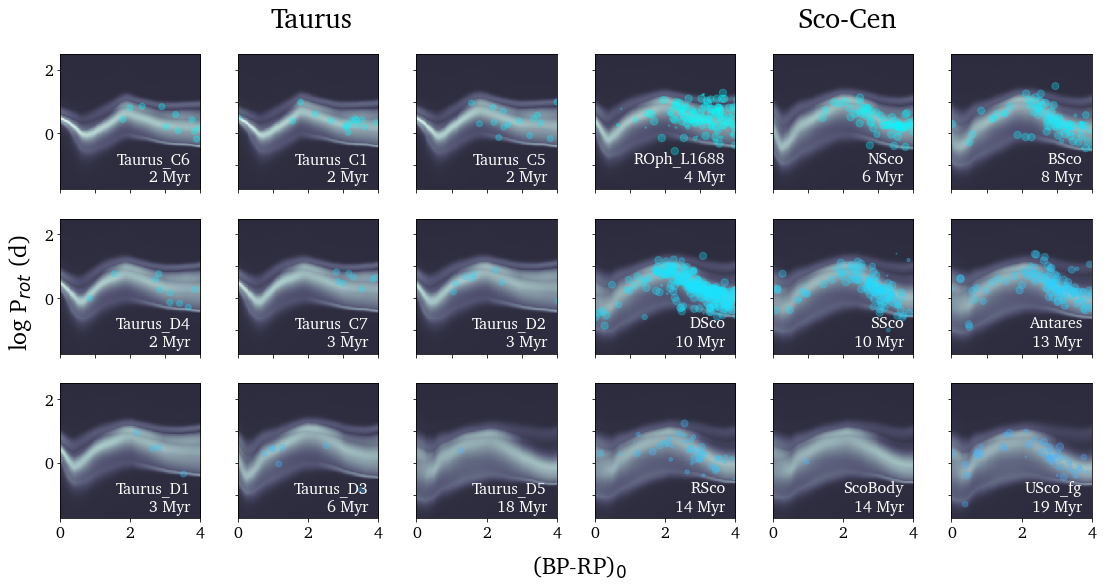}
    \caption{Analagous to Figure \ref{fig: nf_dens_old} but for the subgroups of Taurus and Sco-Cen. The rotational sequences are less defined at these younger ages, as the initial dispersion in rotation period dominates. These subgroups are also much closer in age than the clusters in Figure \ref{fig: nf_dens_old}, and some are poorly populated.}
    \label{fig: nf_dens_young}
\end{figure*}

\begin{figure*}[t]
\centering
\begin{minipage}[t]{0.45\textwidth}
    \includegraphics[width=\textwidth]{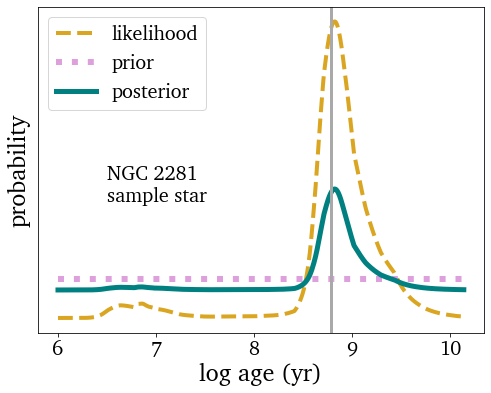}
    \caption{Age probability distributions for a sample star from the NGC 2281 cluster ($\log \tau$ = 9.1, $P_{rot}$ = 10.6 d, $C$ = 0.88). The posterior is clearly likelihood-driven but also smoothed out by the prior, and peaks at an age slightly older than the fiducial cluster age (vertical grey line).}
    \label{fig: ss_sample}
\end{minipage}
\hfill
\begin{minipage}[t]{0.45\textwidth}
    \includegraphics[width=\textwidth]{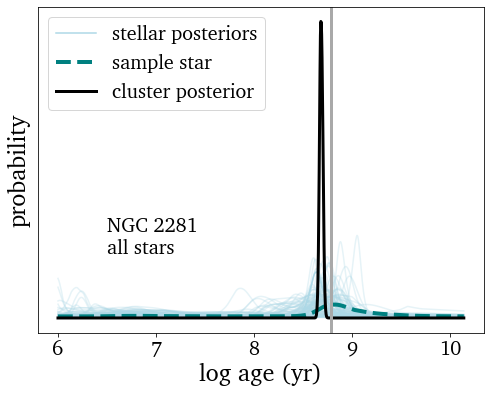}
    \caption{Cluster age posterior for NGC 2281. The stellar posterior from Figure \ref{fig: ss_sample} is plotted here for reference. The fiducial cluster age is plotted as the vertical grey line. Multiplying the individual stellar posteriors together results in a sharply peaked cluster posterior; in this case it slightly underestimates the fiducial cluster age.}
    \label{fig: cl_sample_post}
\end{minipage}
\end{figure*}

\section{ChronoFlow} \label{sec:chronoflow}

Here we introduce \texttt{ChronoFlow}, which models the dependence of rotational evolution on age, observed color, and photometric uncertainty by approximating: $\mathcal{P}(P_{rot}\;|\;C_0,\sigma_{C_0},\tau)$. \texttt{ChronoFlow} implements a conditional normalizing flow (CNF): a neural network-based probabilistic model that excels at modeling complex density distributions.

We trained \texttt{ChronoFlow} to learn $\mathcal{P}(P_{rot}\;|\;C_0,\sigma_{C_0},\tau)$ using the \texttt{zuko}\footnote{\url{https://zuko.readthedocs.io/stable/}} framework, a python-based package that implements CNFs using the \texttt{PyTorch}\footnote{\url{https://pytorch.org/}} deep learning library. In particular, we used a Neural Spline Flow \citep{durkan2019neuralsplineflows} with three transform layers and eight bins, trained for 5,000 epochs. We varied our learning rate using a combination of exponential decay (calibrated to decrease from \texttt{1e-3} to \texttt{5e-7} over 5,000 steps) and a cosine annealing function with a 1,000-epoch period (see \citealt{cosineannealing...loshchilov2017sgdrstochasticgradientdescent}). We chose this setup after experimenting with a variety of model architectures, learning rates, and number of transform layers, as it resulted in the least test loss (i.e., was able to recover our cluster ages with the highest accuracy, using leave-one-out cross validations tests as described in \S\ref{subsec:res_cluster_ages}).

To train \texttt{ChronoFlow}, we used a loss function $\mathcal{L}$ that minimizes the negative log probability of the conditional distribution: $\mathcal{P}(P_{rot}\;|\;C_0,\sigma_{C_0},\tau)$. We further decomposed this probability into two components: the $\mathcal{P}_f$ computed by our flow, and a uniform background probability $\mathcal{P}_b$. $\mathcal{P}_f$ represents the probability density expected from single stars in a cluster following standard spin-down. $\mathcal{P}_b$ is included to account for any non-cluster contaminants. We apply $\mathcal{P}_b$ as:
\begin{equation} \label{eq: bg}
    \mathcal{P}_{b}(P_{rot}) \sim 
    \begin{cases}
    \frac{1}{P_{rot}^{max} - P_{rot}^{min}} & P_{rot}^{min} < P_{rot} < P_{rot}^{max}\\
    0 & \mathrm{otherwise}\\
    \end{cases}
\end{equation}
\noindent where $P_{rot}^{min}$ and $P_{rot}^{max}$ are the limits of log $P_{rot}$ (in days) over which we evaluate the loss; we use -1.75 to 2.5 to accommodate all observational data. Defining $w_f$ as the weighting of our NF probability against our background probability, our loss function is (given data for each $i^{th}$ star out of $n$ total stars):
\begin{equation} \label{eq: loss}
\begin{aligned}
    \mathcal{L} = \Sigma_{i=1}^n [&-\ln(w_{f} \cdot \mathcal{P}_{f}(P_{rot,i}\;|\;C_{0,i},\sigma_{C_0,i},\tau_i)\\&+
    (1-w_{f}) \cdot \mathcal{P}_{b}(P_{rot,i}))]
\end{aligned}
\end{equation}
We also break down our weighting $w_f$ into:
\begin{equation} \label{eq: flow_weighting}
    w_f = (1-p_{out}) \cdot p_{cl}
\end{equation}
Here, $p_{cl}$ is the cluster membership probability for each star (as described in \S\ref{sec: data}), and $p_{out}$ is a universal ``outlier probability'' of 0.05. $p_{out}$ represents the probability of cluster member straying from the single star rotational sequence (e.g., a blue straggler or an unresolved binary). In future phases, applying a dynamic outlier probability per star that uses indications of binarity (such as radial velocity) could further improve this model (see \S\ref{subsec:dis_binarity} for further discussion of binarity).

\section{Results} \label{sec: results}

Figures \ref{fig: nf_dens_old} and \ref{fig: nf_dens_young} present the conditional probability densities calculated by \texttt{ChronoFlow} at the fiducial ages of each of our clusters, including the location of each cluster members in $P_{rot}-C_0$ space for reference. An important note is that the ``high'' probabilities in regions of sparse data at the color extremes are due to the \textit{conditional} nature of our model. Since each subplot represents $p_{f}(P_{rot}\;|\;C_0\!,\sigma_{C_0},\!\tau)$ at a single age, the probability distribution along any vertical line will integrate to 1. Hence, these ``wings'' represent the $p_f(P_{rot})$ distribution that we would expect to see given stars at those ages and colors, not where we would expect to see a high density of stars overall. \textit{This illustrates the importance of applying a conditional normalizing flow to mitigate selection effects,} which otherwise could bias our NF in color space.

To quantitatively measure how well the probability densities from \texttt{ChronoFlow} match observations, a benchmark metric was required. We chose to use the average stellar log probability: \textbf{$p_{*,avg} = \mathcal{P}(P_{rot},C_0\;|\;\sigma_{C_0})$}, i.e. the evidence term in Eqn. \eqref{eq: bayes_with_var}. Clusters with stars concentrated in the high probability regions will have a higher $p_{*,avg}$, and those with more stars outside of those regions will have a lower $p_{*,avg}$. We use the average probability instead of total probability to account for the different sample size of each cluster. With this metric, we conducted two benchmark tests: (i) a cluster-by-cluster comparison, and (ii) an assessment of $p_{*,avg}$ for our observational data against $p_{*,avg}$ of a simulated data set. The details of these tests are described in Appendix \ref{app:model_tests}; overall we find that \textit{\texttt{ChronoFlow} fits observations with high accuracy.}

\begin{figure*}[!htb]
    \centering
    \includegraphics[width=0.98\textwidth]{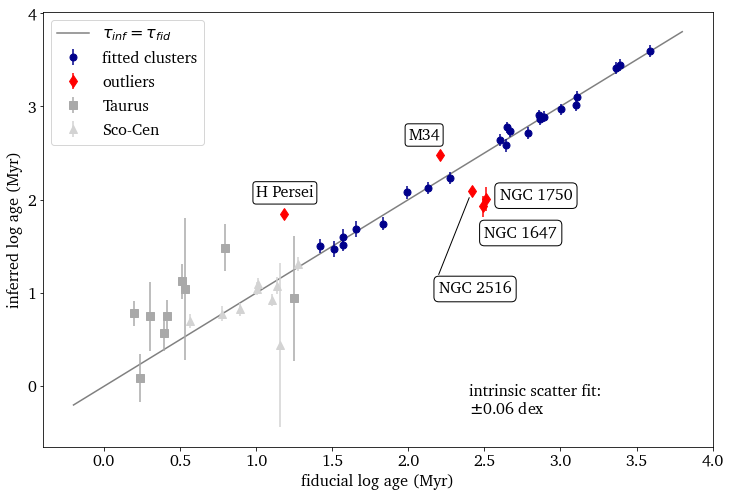}
    \caption{Inferred cluster ages from LOOCV compared to the fiducial values from literature. Subgroups of the Taurus and Sco-Cen regions are plotted in grey, and the five anomalous clusters are plotted in red. The best fit line is $\tau_{\mathrm{inf}} = (1.00\pm0.02)\cdot\tau_{\mathrm{fid}} - (0.00\pm0.06)$, with an intrinsic scatter of 0.06 in $\tau_{\mathrm{inf}}$. This indicates that there is \textit{no systematic offset} in $\tau_{\mathrm{inf}}$ resulting from the model itself. Error bars include both the statistical uncertainty for each cluster estimate and the intrinsic scatter.}
    \label{fig: loocv}
\end{figure*}

\begin{figure*}
    \centering
    \includegraphics[width=0.98\textwidth]{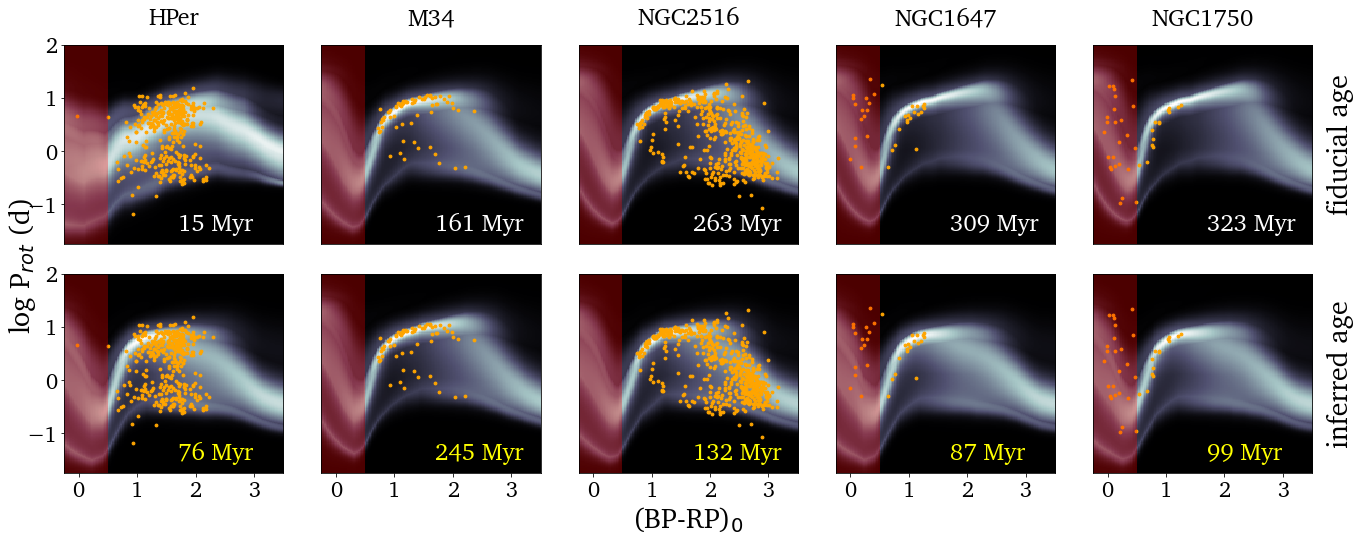}
    \caption{A comparison of the five anomalous clusters from Figure \ref{fig: loocv}. Observed data are plotted over our model's probability density at both fiducial (top) and inferred (bottom) cluster ages. The same default parameters for $\sigma_{C_0}$ and $p_{cl}$ were used as in Figure \ref{fig: nf_dens_old}. We exclude stars having $(BP-RP)_0 <$ 0.5 from our age inference, and have shaded this region in red. The \texttt{ChronoFlow} estimates for M34 and NGC 2516 are consistent with literature age estimates of M34 \citep{GaiaCollab...2018AA...616A..10G,Ianna...M34...1993AJ....105..209I,Meibom...M34...2011ApJ...733..115M} and NGC 2516 \citep{Meynet...OCages...1993A&AS...98..477M,GodoyRivera...2021...rotcatalogue,Sung...NGC2516...2002AJ....123..290S,Richer...OCages...2021ApJ...912..165R,Bouma...NGC2516...2021AJ....162..197B}.}
    \label{fig:anomalous_clusters}
\end{figure*}

\subsection{Stellar and cluster age recovery} \label{subsec:res_cluster_ages}

Evaluating Eq. \eqref{eq: bayes_with_var_expanded} over a grid of ages, for a constant set of observed parameters, is straightforward, and this generates a posterior age probability for each star. Figure \ref{fig: ss_sample} presents an example of this. As expected, given the uniform prior, the posterior follows the likelihood closely.

To obtain the posterior age estimation for a cluster $\mathcal{
P}(\tau_{cl})$, we take the product of the individual stellar likelihood functions for all $n_{cl}$ stars in the cluster, and multiply that by our age prior $\mathcal{P}(\tau\;|\;\sigma_{C_0})$. Since the stellar likelihood consists of both the $\mathcal{P}(P_{rot}\;|\;C_0,\sigma_{C_0},\tau)$ and $\mathcal{P}(C_0\;|\;\tau,\sigma_{C_0})$ terms, we calculate each cluster posterior using:
\begin{equation} \label{eq: cluster_post_1}
\begin{aligned}
    \mathcal{P}(\tau_{cl}) = \mathcal{P}(&\tau\;|\;\sigma_{C_0}) \\ \cdot \prod_{i = 1}^{n_{cl}} [ \mathcal{P} &(P_{rot,i}\;|\;C_{0,i},\sigma_{C_0,i},\tau) \\ & \cdot \mathcal{P}(C_{0,i}\;|\;\tau,\sigma_{C_0,i})]
\end{aligned}
\end{equation}
\noindent We have ignored the evidence term here because we re-normalize the cluster posteriors to integrate to 1.\footnote{Functionally, since we have applied a uniform age prior and are normalizing $\mathcal{P}(\tau_{cl})$, we can instead calculate the cluster posterior as the product of all stellar posteriors in the cluster.} Figure \ref{fig: cl_sample_post} illustrates the cluster posterior calculation for NGC 2281 as an example.

When calculating these cluster posteriors, we exclude stars having $C_0 >$ 0.5 because the stability of our model suffers in this regime. The reason for this behavior is likely twofold: (i) we have sparse sampling; and (ii) those stars may predominantly be above the Kraft break (\citealt{Kraft...spindown...1967ApJ...150..551K}), meaning they have radiative envelopes and do not undergo magnetic braking. \citet{Beyer...KraftBreak...2024ApJ...973...28B} identify the Kraft break at $(BP-RP)_0 \approx 0.54-0.60$, indicating that $C_0 = (BP-RP)_0 = 0.5$ is a reasonably conservative cut that should not exclude any useful data.

As a cluster age recovery exercise, we applied our trained model to Eqn.~\eqref{eq: bayes_with_var_expanded} using leave-one-out cross validation (LOOCV), where each cluster age was inferred using a model that \textit{excluded} the cluster from training. Figure \ref{fig: loocv} presents these ages compared to the fiducial literature values. There is very good agreement, aside from five outliers. Excluding those five clusters, we fit a line to the $\tau_{\mathrm{inf}}$ vs. $\tau_{\mathrm{fid}}$ relationship using \texttt{emcee}\footnote{\url{https://emcee.readthedocs.io/en/stable/}} \citep{emcee...2013PASP..125..306F}. We included a free parameter for intrinsic scatter, which we found to be 0.06 dex. Given that the statistical uncertainty in the individual cluster posteriors (median of 0.016 dex) is therefore dominated by the intrinsice scatter, we conclude that \textit{\texttt{ChronoFlow} can infer cluster ages to an overall statistical uncertainty of 0.06 dex, or $\approx$15\%}. Since these results are from our LOOCV exercise, each estimate is \textit{simulating a scenario where our model is encountering new data}, so this is the level of accuracy we would expect from \texttt{ChronoFlow} applied to new populations. We do not expect precise age estimations for the subgroups of Taurus and Sco-Cen, since they generally have limited data and stars at such young ages have large $P_{rot} $dispersion. We see a clear systematic where Taurus ages are overestimated, but nevertheless \texttt{ChronoFlow} recovers the young sungroup ages to within uncertainties reasonably well.

\subsubsection{Outliers}
\label{subsubsec:res_anomalous_clusters}

Here we discuss possible reasons for the poor match between observations and \texttt{ChronoFlow} for the aforementioned five anomalous clusters.

H Persei has an abundance of fast rotators relative to other clusters of similar ages, and a more dispersed slowly rotating sequence. \citet{Moraux...gyro...HPer...2013AA...560A..13M} note that the fast rotators may represent a population of tidally locked close binary systems. \citet{Getman...HPer...2023ApJ...952...63G} suggest that the $P_{rot}$ measurements may also be contaminated by aliases from solar and lunar cycles. This possible overabundance of fast rotators does not intuitively explain why our model \textit{over}-estimates the age of H Persei, but these factors suggest that the observations we are using may not be an accurate representation of the $P_{rot}$ evolution of single stars at that age.

Our inferred age for M34 of $245^{+40}_{-34}$ Myr does not match the ages from \citealt{CantatGaudin...catalogue...2020AA...640A...1C} (132 Myr) or \citealt{Bossini...Catalogue...2019AA...623A.108B} (126 Myr) well, but is more consistent with \citealt{GaiaCollab...2018AA...616A..10G} (251 Myr). There is also significant variation in other literature isochronal ages, ranging from e.g. $132\pm63$ Myr (\citealt{Richer...OCages...2021ApJ...912..165R}) to 250 Myr (\citealt{Ianna...M34...1993AJ....105..209I}). Notably, \citet{James...M34...2010A&A...515A.100J} and \citet{Meibom...M34...2011ApJ...733..115M} inferred rotational ages of $193\pm9$ Myr and 240 Myr respectively, so our age is consistent with this larger set of literature estimates.

NGC 2516 is another cluster for which literature values vary significantly. The average isochronal age from our main catalogs is 263 Myr, but the next oldest literature age is 200 Myr (\citealt{Richer...OCages...2021ApJ...912..165R}), and we find four more even younger ages (the youngest being $102\pm15$ Myr from \citealt{Li...NGC2516...2024A&A...686A.142L}). Our estimate of $132^{+20}_{-18}$ Myr is consistent with these younger literature values , and also consistent with the rotational ages estimated by \citealt{GodoyRivera...2021...rotcatalogue} (150 Myr).

We only have 14 stars in NGC 1750 having $C_0 >$ 0.5, and the reddest has $C_0$ = 1.25, so the small sample size and lack of color coverage may hinder our age inference for this cluster. NGC 1750 is also less studied than some of the other clusters, with only two isochronal ages we could find outside of our main catalogs. These two estimates are younger than the fiducial age of 323 Myr ($257 \pm 123$ Myr from \citealt{Richer...OCages...2021ApJ...912..165R} and $200 \pm 50$ Myr from \citealt{GaladiEnriquez...NGC1750...1998A&A...333..471G}), but still not consistent with our inferred age of $99^{+32}_{-21}$ Myr.

Small sample size and color coverage may also affect NGC 1647, for which we have 19 total stars with $C_0 > 0.5$ and a maximum $C_0$ of 1.26. More isochronal age estimates are available in literature for this cluster, and similar to NGC 1750 they are on average younger than those from our main catalogs (fiducial age of 309 Myr), with five ages ranging from 110 \citep{Turner...NGC1647...1992AJ....104.1865T} to 200 ($\pm50$; \citealt{Richer...OCages...2021ApJ...912..165R}) Myr. Again, these are closer to (but still inconsistent with) our inferred age of $87^{+25}_{-18}$ Myr.

Figure \ref{fig:anomalous_clusters} presents the observed data for those clusters plotted against our flow results at: (i) the fiducial literature age, and (ii) the age that the final version of \texttt{ChronoFlow} (from which we exclude these five clusters during training) infers. It is clear that H Persei seems to be an outlier than our model is unable to characterize well. Our model fits NGC 1647 and NGC 1750 better at our inferred ages than the fiducial ages, so we present that as tentative evidence that these clusters are on the younger end of literature estimates, however we do not claim these age estimates to be accurate due to the lack of sample size and color coverage. In contrast, M34 and NGC 2516 are well sampled, and our inferred ages fit the observations, so we present these as well-motivated rotational ages:

\begin{itemize}[itemsep=-2pt]
    \item \textbf{M34:} $245^{+40}_{-34}$ Myr
    \item \textbf{NGC 2516:} $132^{+20}_{-18}$ Myr
\end{itemize}

\noindent which support literature estimates of M34 \citep{GaiaCollab...2018AA...616A..10G,Ianna...M34...1993AJ....105..209I,Meibom...M34...2011ApJ...733..115M} and NGC 2516 \citep{Meynet...OCages...1993A&AS...98..477M,GodoyRivera...2021...rotcatalogue,Sung...NGC2516...2002AJ....123..290S,Richer...OCages...2021ApJ...912..165R,Bouma...NGC2516...2021AJ....162..197B}. Our reported uncertainties include both the statistical uncertainty for each cluster and the intrinsic scatter of our model.

\subsection{Age recovery for single stars}
\label{subsec:res_single_star_ages}

\begin{figure*}
    \centering
    \includegraphics[width=0.93\textwidth]{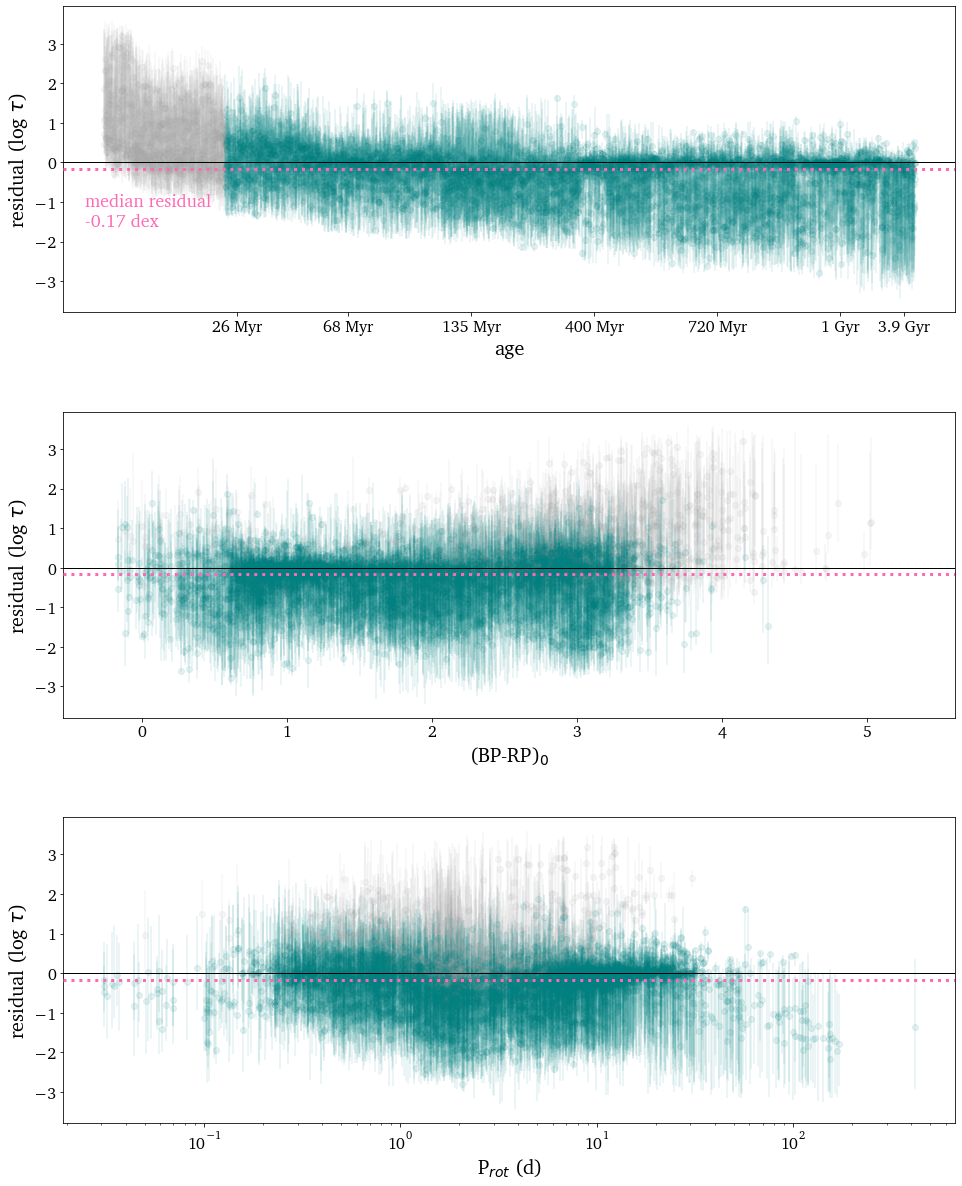}
    \caption{Individual stellar age residuals from posterior estimates. Grey stars are members of the Taurus and Sco-Cen associations, and all other clusters are plotted in teal. In the top panel, stars are uniformly distributed along the age axis in order of increasing age, instead of plotted directly against age, for ease of visualization. The error bars encapsulate the 16th to 84th percentiles of every stellar posterior estimate. The black solid line is plotted at a residual of 0, and the pink dotted line is plotted at the median residual over our entire catalog (excluding the Taurus and Sco-Cen members), which is -0.17.}
    \label{fig:single_star_residuals}
\end{figure*}

\begin{figure}
    \centering
    \includegraphics[width=0.45\textwidth]{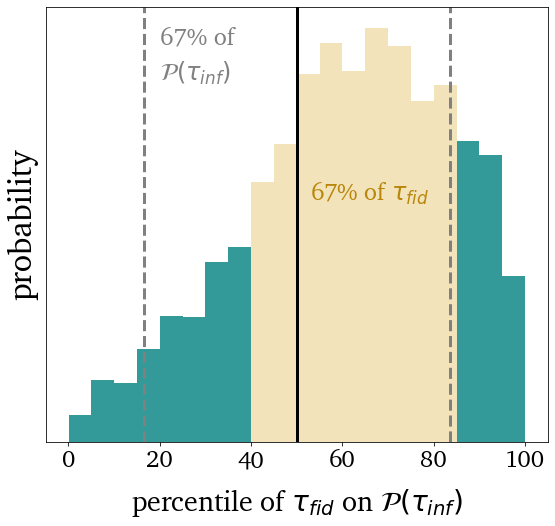}
    \caption{Distribution of the percentiles at which $\tau_{fid}$ lies in each $\mathcal{P}(\tau_{inf})$ posterior. The posteriors systematically underestimate the fiducial stellar ages, but overestimate the uncertainty. The $\pm1\sigma$ error in the posteriors is bounded by the dashed grey lines (the solid black line is the median), and the $\pm1\sigma$ distribution of fiducial literature ages $\tau_{fit}$ is shown as the light yellow region.}
    \label{fig:single_star_err_hist}
\end{figure}

Our focus to this point has been on characterizing the rotational evolution of a population of stars, but now we examine individual stars. As a standalone test, we applied \texttt{ChronoFlow} to the Sun using $P_{rot}$ = 26.09 days and $(BP-RP)_0$ = 0.82 (from \citealt{Casagrande...SunBPRP...2018MNRAS.479L.102C}), and calculated its age to be $5.1^{+1.7}_{-1.4}$ Gyr. This is consistent at $1\sigma$ with the true age of 4.6 Gyr.

More comprehensively, we conducted another version of LOOCV, but using batches of stars drawn from all clusters instead of testing a single cluster at a time. We divided our data catalog (excluding H Persei, M34, NGC 2516, NGC 1647, and NGC 1750) into 20 batches, and trained 19 different models that each included 19/20 batches (95\% of our data). Clusters were distributed as equally as possible between batches to ensure maximal age coverage in each batch. We then used each model to calculate stellar age posteriors for the 5\% of stars that were left out of training.

\begin{figure}
    \centering
    \includegraphics[width=0.48\textwidth]{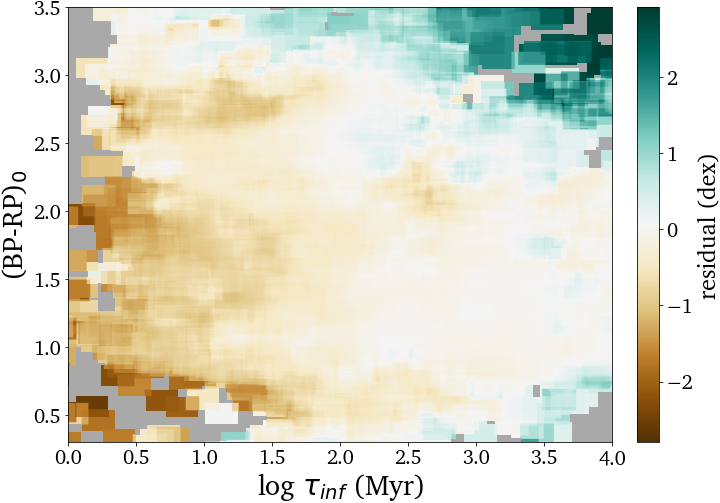}
    \caption{Average residuals as a function of color and inferred age. Data are averaged over bins of width 0.2 in both dimensions. Grey shaded regions are bins with no data.}
    \label{fig:avg_res_matrix}
\end{figure}

\begin{figure}
    \centering
    \includegraphics[width=0.48\textwidth]{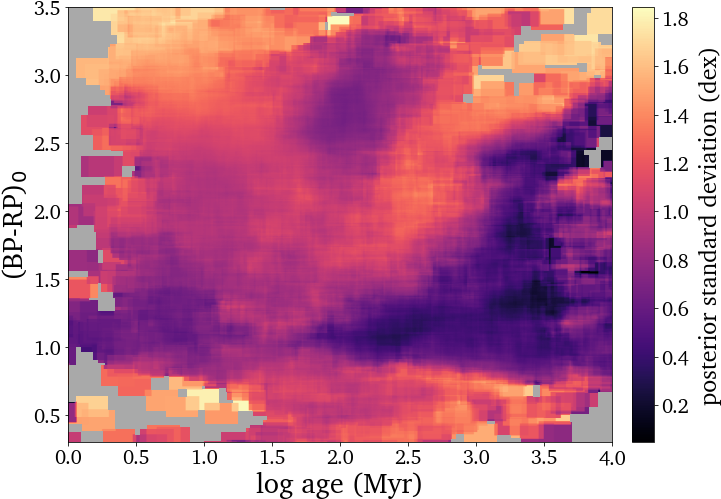}
    \caption{Average standard deviation of individual stellar posteriors as a function of color and inferred age. Data are averaged over bins of width 0.2 in both dimensions. Grey shaded regions are bins with no data.}
    \label{fig:avg_prec_matrix}
\end{figure}

Figure \ref{fig:single_star_residuals} presents these results. The top panel shows the residuals as a function of stellar age, the middle panel plots the residuals against $C_0$, and the bottom panel against measured $P_{rot}$. Members of Taurus and Sco-Cen are plotted in grey; we do not expect this model to reliably recover the ages of such young stars.

There is a clear trend at older ages: a dense population of stars that the model estimates well, and a dispersed tail that the model underestimates, which is reflected in the median residual value of -0.17 dex. These ``underestimated'' stars are those in the wide distribution of fast rotators in their cluster, as confirmed from the lower panel where these stars generally have an intermediate rotation period of $\approx$ 11-17 days. The model interprets these as being on a slowly rotating sequence at a younger age. 

Overall, 79\% of stars have a fiducial age that lies in the 1$\sigma$ credibility interval of our posterior age estimates, so we are \textit{overestimating} the uncertainty on these posteriors. Figure \ref{fig:single_star_err_hist} illustrates this. The median $1\sigma$ uncertainty on the residuals is [-0.78,0.64] dex, but we also note that both the stellar residuals and stellar posterior precisions exhibit systematic trends with color and inferred age. Figure \ref{fig:avg_res_matrix} illustrates that \texttt{ChronoFlow} is relatively unbiased for stars within a $(BP-RP)_0$ range of [1.0,2.5] for which it infers ages of $\gtrsim 60$ Myr. For stars with an inferred age younger than that, \texttt{ChronoFlow} is likely underestimating the true age, especially for the bluest stars. We also see that for stars having $(BP-RP)_0 \gtrsim 2.5$, inferred ages of $> 1$ Gyr may be significant overestimations. Figure \ref{fig:avg_prec_matrix} shows that the regime of least bias described above generally corresponds with the regime of highest stellar posterior precision. Precision notably suffers for stars having $(BP-RP)_0 \gtrsim 3.0$ or $(BP-RP)_0 \lesssim 0.75$.

Therefore, while we summarize the statistical uncertainty for individual stars as $\approx$ 0.7 dex, these systematics should be considered when using \texttt{ChronoFlow} for \textit{individual} stellar age estimates.

It is also important to note that there is almost certainly some level of contamination in our cluster sample, and therefore for non-members we are incorrectly using the cluster ages as the true ages of these stars. However, this effect is difficult to quantify since by definition the contaminants have not been identified.

\subsection{Comparison to literature models}
\label{subsec:model_comp}

Here we discuss the results and performance of \texttt{ChronoFlow} relative to the most comparable literature models: \texttt{GPgyro}\footnote{\url{https://github.com/lyx12311/GPgyro}} by \citet{Lu...Day&Age...2024AJ....167..159L} and \texttt{gyro-interp}\footnote{\url{https://github.com/lgbouma/gyro-interp}} by \citet{Bouma...gyrointerp...2023ApJ...947L...3B}. We performed two tests: (i) a cluster age recovery exercise, and (ii) a residual analysis analogous to \S\ref{subsec:res_single_star_ages}.

Since the other two models use $T_{\text{eff}}$ instead of $(BP-RP)_0$, we restricted the stars used in these tests to the subset of our catalog having a measured Gaia DR3 \texttt{teff\_gspphot}. Additionally, we follow the guidance of the other models' effective parameter space to perform additional cuts. For \texttt{GPgyro}, we apply:

\begin{itemize}[itemsep=0pt]
    \item 3000 $\leq$ \texttt{teff\_gspphot} $\leq$ 7000
    \item Absolute Gaia DR3 magnitude $M_G < 4.2$ (derived from DR3 parallax)
    \item Stellar age (from cluster membership) $\geq$ 670 Myr
\end{itemize}

\noindent It is also worth noting that \texttt{GPgyro} is not expected to perform as well for stars $\lesssim$ 1.5 Gyr, however we still include these for completeness. For \texttt{gyro-interp}, we apply:

\begin{itemize}[itemsep=0pt]
    \item 3800 $\leq$ \texttt{teff\_gspphot} $\leq$ 6200
    \item Stellar age (from cluster membership) $\geq$ 80 Myr
\end{itemize}

\begin{figure*}
    \centering
    \includegraphics[width=0.9\textwidth]{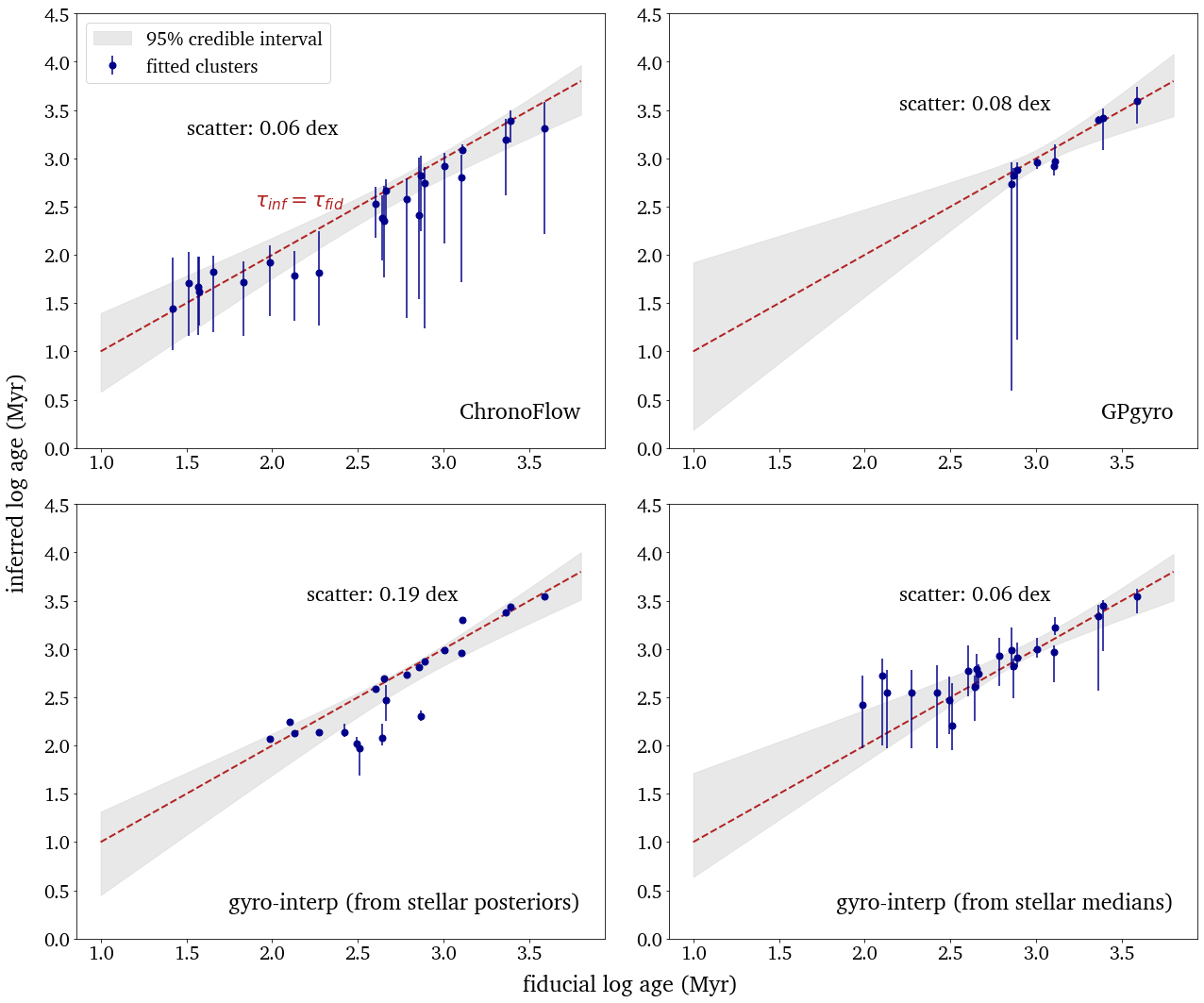}
    \caption{Cluster age inference results from each model when applied to the subset of our data catalog in the models' effective regimes. The outlier clusters described in \S\ref{subsubsec:res_anomalous_clusters} are included. Only stars with a Gaia DR3 \texttt{teff\_gspphot} measurement were included in this exercise, hence the discrepancy with Figure \ref{fig: loocv}. \texttt{ChronoFlow} was evaluated using LOOCV, while we used the published version of the other models. We shade the 95\% credible interval of a linear fit to $\tau_{\text{inf}}$ vs. $\tau_{\text{fid}}$ for each model. Overall, the accuracy of each model is comparable, however \texttt{ChronoFlow} maintains its accuracy to younger ages than the other models.}
    \label{fig:model_comp_loocv}
\end{figure*}

For the cluster age recovery exercise, we first computed the stellar posteriors with each model. Since \texttt{GPgyro} outputs stellar median ages with uncertainties, and not an age probability distribution, we computed the cluster estimates as the distribution of the median inferred stellar ages. While \texttt{gyro-interp} outputs complete stellar age posteriors, since it does not include an outlier model that forces a non-zero probability at every age, taking the product of stellar posteriors directly often results in a null cluster posterior. So, we calculated cluster age probabilities in two ways. The first method was the same as for \texttt{GPgyro}. In the second, we manually added the same 5\% outlier probability that we include in \texttt{ChronoFlow} to the individual \texttt{gyro-interp} stellar posteriors, and then multiplied them to get the cluster posteriors. Both results are presented in Figure \ref{fig:model_comp_loocv}.

It should also be noted that the datasets that we have used to compare the models likely introduce several systematics in the cluster age recovery test. First, the other models were not trained on Gaia DR3 $T_{\text{eff}}$, so age inference using that temperature prescription may decrease the performance. However, we considered modeling $T_{\text{eff}}$ for our datalog to be outside the scope of this study. A competing effect is that we used LOOCV when testing the cluster age recovery of \texttt{ChronoFlow} in this context, but our data catalog includes stars that were used to train the other models, so we expect their performance to be better given that the test data is not entirely new. We have not attempted to quantify these effects, but to first order assume that they cancel out and ignore them in our evaluation.

The results of the age inference test (Figure: \ref{fig:model_comp_loocv}) demonstrate that in the region of parameter space where the models overlap, the performance across models is comparable. The benefit of \texttt{ChronoFlow} is that it extends to younger ages than the other two. We also see that training \texttt{ChronoFlow} on this subset of data decreases performance significantly compared to the model trained on our entire data catalog, indicating that the extent of our training catalog is an an important factor in performance, and perhaps that additional systematics are introduced by only considering stars with $T_\text{eff}$ measurements in Gaia DR3.

\begin{figure*}
    \centering
    \includegraphics[width=0.98\textwidth]{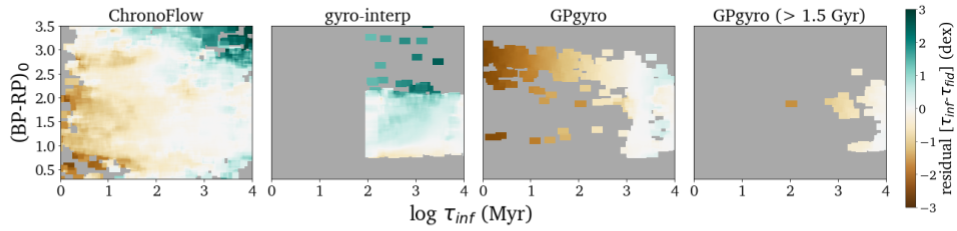}
    \caption{A comparison of the stellar residuals between \texttt{ChronoFlow}, \texttt{gyro-interp}, and \texttt{GPgyro} in $\tau_{\text{inf}}-(BP-RP)_0$ space, averaged across bins of 0.2 in both dimensions. The residual is the inferred age from each model minus the fiducial age according to our catalog. The left panel is a copy of Figure \ref{fig:avg_res_matrix}. It is evident that \texttt{ChronoFlow} has more complete coverage over the parameter space at open cluster ages than the other models, and it exhibits less to equivalent systematic uncertainty compared to the other models in the regions of overlap.}
    \label{fig:model_comp_residuals}
\end{figure*}

Figure \ref{fig:model_comp_residuals} presents the residual analysis for all three models (the left panel is a copy of Figure \ref{fig:avg_res_matrix}). It is apparent that the applicable parameter space of \texttt{ChronoFlow} extends younger than that of the other two models, both overall and when considering regions of low bias. There is also wider coverage across the color space, although the other models use $T_{\text{eff}}$ in place of color so that comparison is more difficult. In the regime of overlap between \texttt{ChronoFLow} and \texttt{GPgyro}, the models are comparable and both relatively unbiased. \texttt{GPgyro} is more biased when applied at lower inferred ages, however applying the stricter cut of $> 1.5 Gyr$ to the test data set removes most of that bias (although also reduces the parameter space; see the farthest right panel). Where \texttt{ChronoFlow} and \texttt{gyro-interp} overlap, both models perform well, however the residuals are systematically slightly higher in \texttt{gyro-interp}. This indicates that \texttt{ChronoFlow} as a model may impose less structure across the parameter space, and therefore demonstrate better flexibility in age inference.

\begin{figure}
    \centering
    \includegraphics[width=0.48\textwidth]{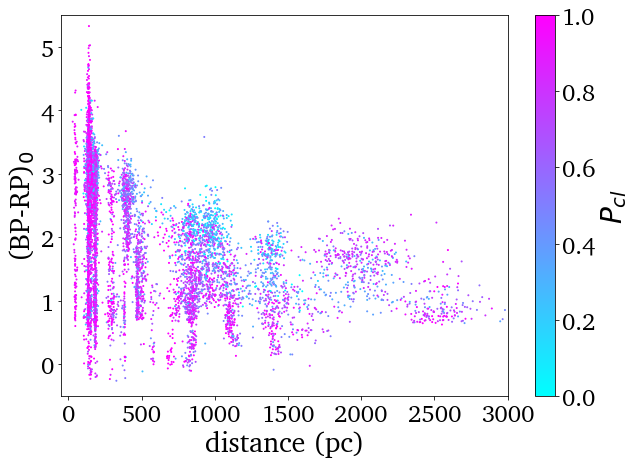}
    \caption{HDBScan Cluster membership probabilities as a function of color and distance. As expected, there is weaker confidence in stars that are redder and farther away, since their astrometry is less precise.}
    \label{fig:Pclmem_systematic}
\end{figure}

\section{Systematic Uncertainties} \label{sec: discussion}

\noindent In \S\ref{sec: results} we evaluated the statistical uncertainty associated with our age inference. However, there are additional systematic uncertainties that are important considerations for any gyrochronology model. We tested these to understand how important they are to our overall error budget:

\begin{itemize}
    \item \textbf{De-reddening}. The choice of dustmap used to derive stellar extinctions directly affects photometric color. We analyze this in two ways, by: (i) characterizing the inherent deviation caused by using different dustmaps for both training and inference, and (ii) predicting ages using photometry from a different dustmap than was used to calibrate the model. The impact of the former is relatively small, but the impact of the latter can be significant. Details are provided in Appendix \ref{subsec:sys_dustmap_comparison}.
    \item \textbf{Cluster membership cuts}. We analyze the impact of cluster membership cuts by comparing predictions for three clusters derived from three different data catalogs for each cluster: the Pleiades, Praesepe, and NGC 6811. Details are provided in Appendix \ref{subsec:disc_sys_dc_comparison}. The impact of this was negligible.
    \item \textbf{Source survey}. We test for hidden systematics by evaluating our age inference both with and without ground based data in our training catalog. Including those data improves our age recovery and does not introduce additional systematic uncertainty, indicating that they add value to our catalog. Details are provided in Appendix \ref{subsec:disc_spaceonly}.
    \item \textbf{Calibration ages}. Since cluster ages are inherently model-dependent, the choice of model calibration ages affects age inference. We examine this impact using a variety of training ages, and find this to be a significant systematic.
    \item \textbf{Bias in $p_{cl}$}. Since HDBScan cluster membership probabilities are reliant on Gaia astrometry, they are fundamentally biased lower towards fainter (i.e. redder and more distant) stars which have larger astrometric errors. Figure \ref{fig:Pclmem_systematic} illustrates this. Additionally, since HDBScan results are heavily dependent on the hyperparameters, it is important to quantify how that impacts age inference.
\end{itemize}

\begin{table*}
\centering
\renewcommand{\arraystretch}{1.3}
\begin{tabular}{|c|p{10cm}|c|}
\hline
\textbf{Metric} & \textbf{Description} & \textbf{Value (dex)} \\
\hline
\hline
$\sigma_{\mathrm{A,1}}$ & Systematic uncertainty due to choice of dust map in de-reddening. & 0.01 \\
\hline
$\sigma_{\mathrm{A,2}}$ & Systematic uncertainty resulting from different de-reddening processes between calibration photometry and prediction photometry. & $\lesssim$ 0.04 \\
\hline
$\sigma_{\mathrm{cl}}$ & Systematic uncertainty caused by varying cluster membership criteria.  & 0.01 \\
\hline
$\sigma_{\mathrm{\tau,cal}}$ & Systematic uncertainty due to varying the model calibration age. & $\approx$ 0.04 \\
\hline

\end{tabular}
\caption{Summary of our systematic test results. The dominant sources of uncertainty come from using different extinctions for training and inference, and the model-dependency of calibration ages themselves. Added in quadrature, these result in a total systematic uncertainty of $\approx$ 0.06 dex.}
\label{tab:systematics}
\end{table*}

\noindent We summarize our results for these systematics in Table \ref{tab:systematics}. The results of the space vs. ground based survey test indicated that there was no quantifiable systematic, so that is not included. 

While the effect of the bias in $p_{cl}$ with distance and color cannot be reliably quantified, qualitatively this results in wider distributions of $\mathcal{P}(P_{rot}\;|\;C_0,\sigma_{C_0},\tau)$ in that regime, due to the increased weighting of the uniform background probability $\mathcal{P}_b$. We test the influence of $p_{cl}$ on our model by comparing LOOCV results using HDBScan probabilities against LOOCV results using $p_{cl}$ from the source catalogs. The difference in cluster posteriors, outliers, and intrinsic scatter is negligible, indicating that this systematic is insignificant.

\subsection{Impact of binarity}
\label{subsec:dis_binarity} 

We have treated potential binary contamination simplistically in this work, by first applying cuts in an attempt to restrict our training data to only single stars (using Gaia DR3 \texttt{RUWE} and literature flags), and then including a uniform outlier component in our model to account for binaries that passed those cuts. We recognize that this uniform outlier is naive; there are additional data such as radial velocities that we could use to better inform the outlier probability for each star, and it would also be possible to model a \textit{binary} evolution track in \texttt{ChronoFlow} in addition to the single star track. This could help account for systematic rotational trends, such as the overdensity of tidally synchronized binaries among fast rotators in the Kepler field identified by \citet{Simonian...BinaryRotation...2019ApJ...871..174S}, and the abundance of binaries in the fast rotating sequence of Blanco 1 and the Pleiades found by \citet{Gillen...Blanco1...2020MNRAS.492.1008G}. Intentionally modeling binaries would also be preferable to attempting to remove them entirely from our model, since (i) there will likely always be some level of contamination, and (ii) binaries make up a significant portion of the stellar population. The behavior of binary contaminants may therefore be an additional unmitigated source of systematics in our current model, but we leave a more comprehensive treatment for future work.

\subsection{Other considerations}
\label{subsec:dis_other_considerations}

While we have mitigated the effects of detection bias in color and age by using a \textit{conditional} normalizing flow, selection effects in $P_{rot}$ still exist. $P_{rot}$ measurements are naturally biased towards stars with shorter periods (since these rotate more times per observational baseline), and greater periodic variability amplitude (which results in a stronger signal). Overall, the concentration of cluster members at reliably measurable periods indicates that we do not expect there to be a significant population of stars at undetectable periods, at least up to the age of M67 where we do see some very long rotators. However, older fainter stars (or high mass stars) with weak magnetic activity may have low rotational modulation amplitudes due to a lack of star spots, and young stars may have extreme short term activity-induced variability that obscures the rotational signal. In both cases, our ability to measure a rotation period is hindered. We leave that as a caveat to this work, but it also begs the question of whether including activity-based metrics in a model such as \texttt{ChronoFlow} can improve age constraints; this is a topic we intend to investigate in future work.

Another attribute that affects rotational evolution is metallicity, as its influence on opacity affects the depth of the surface convective zone and therefore the amount of magnetic braking (e.g. \citealt{Claytor...RotationalAges...2020ApJ...888...43C, Lu...Day&Age...2024AJ....167..159L}). It would be straightforward to include metallicity as a training parameter in our model, however most of the open clusters are approximately solar metallicity since they are nearby. Given this lack of parameter space coverage combined with the uncertainties on measurements, we do not expect our model to learn anything meaningful from including metallicity. Should significant rotation data become available for stars with super- or sub- solar metallicity, this would be worth investigating.

In future work, it is important to consider how to optimize this framework for the lowest mass (e.g. K and M) stars. In that regime, the Gaia $(BP-RP)_0$ is not as descriptive given the low flux in the BP band. This is seen for M67 in Figure \ref{fig: nf_dens_old}; there appear to be many stars above the slow rotator sequence in a narrow color range, but when \citet{Dungee...gyro.M67...2022ApJ...938..118D} plot those rotation periods as a function of stellar effective temperature ($T_{\text{eff}}$) the sequence is smooth. We would expect the Gaia $(G-RP)_0$ color to be more descriptive than $(BP-RP)_0$, so including both as model parameters could improve performance.

\section{Applications}
\label{sec:applications}

\subsection{Ages for coeval populations} \label{subsec:res_new_cluster_ages}

\begin{figure}[!htb]
    \centering
    \plotone{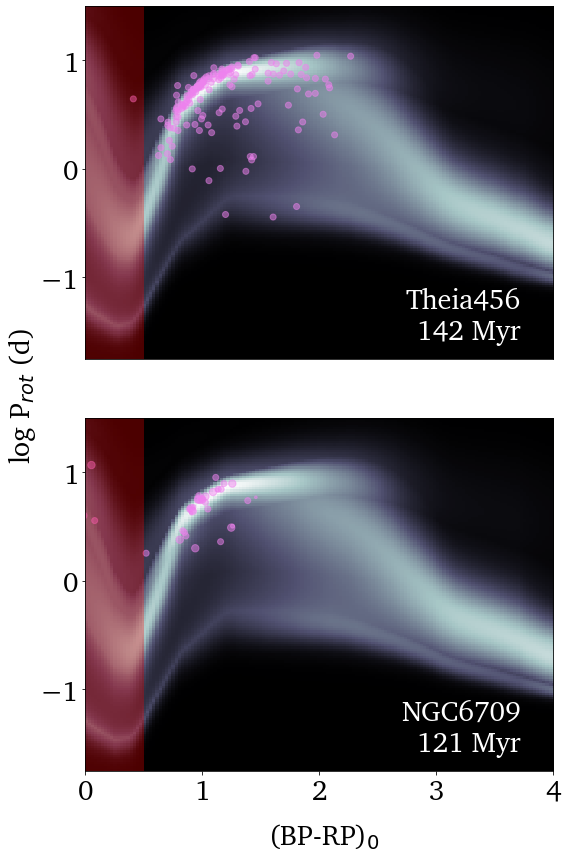}
    \caption{A visualization analogous to Figure \ref{fig: nf_dens_old} showing the probability densities generated by our model at our inferred ages for the Theia 456 stellar stream and the NGC 6709 open cluster, along with the observational data. The stars in NGC 6709 are sized according to cluster membership probability; those values were not provided in the Theia 456 catalog so all stars are sized equally (reflective of our fiducial cluster membership probability of 0.9). We exclude stars having $(BP-RP)_0 < 0.5$ from our age inference, and have shaded this region in red.}
     \label{fig:new_cluster_densities}
\end{figure}

In addition to generating rotational age estimates for M34, NGC 2516, NGC 1647, and NGC 1750 as described in \S\ref{subsec:res_cluster_ages}, we applied our model to two systems that were not considered in our data catalog: the NGC 6709 open cluster (using rotation periods from \citealt{ColeKodikara...NGC6709...2023A&A...673A.119C}), and the Theia 456 stellar stream (using rotation periods from \citealt{Andrews...Theia456...2022AJ....163..275A}). Gaia EDR3 IDs were provided in both cases, so we calculated the de-reddened photometric uncertainty using the process described in \S\ref{subsec: photometry} using the \cite{Edenhofer2023...dustmap...2023arXiv230801295E} dust map. We retained 161 of the 171 Theia 456 stars having reported rotation periods; one was excluded because the ID provided did not match any records in Gaia DR3 or EDR3, eight more were excluded because they have Gaia RUWE values $>$ 1.4, and one more was excluded because it has $C_0 < 0.5$. For NGC 6709, we retained 34 out of the 48 stars with reported rotation periods. 11 were excluded because they were flagged as binaries by \cite{ColeKodikara...NGC6709...2023A&A...673A.119C}, and none of the 37 remaining had a Gaia RUWE $>$ 1.4, but three also had $C_0 < 0.5$, which we excluded. We then calculated the system age posteriors for each, following the methodology described in \S\ref{subsec:res_cluster_ages} (including the intrinsic scatter in our uncertainties).

\begin{figure*}
    \centering
    \includegraphics[width=0.95\textwidth]{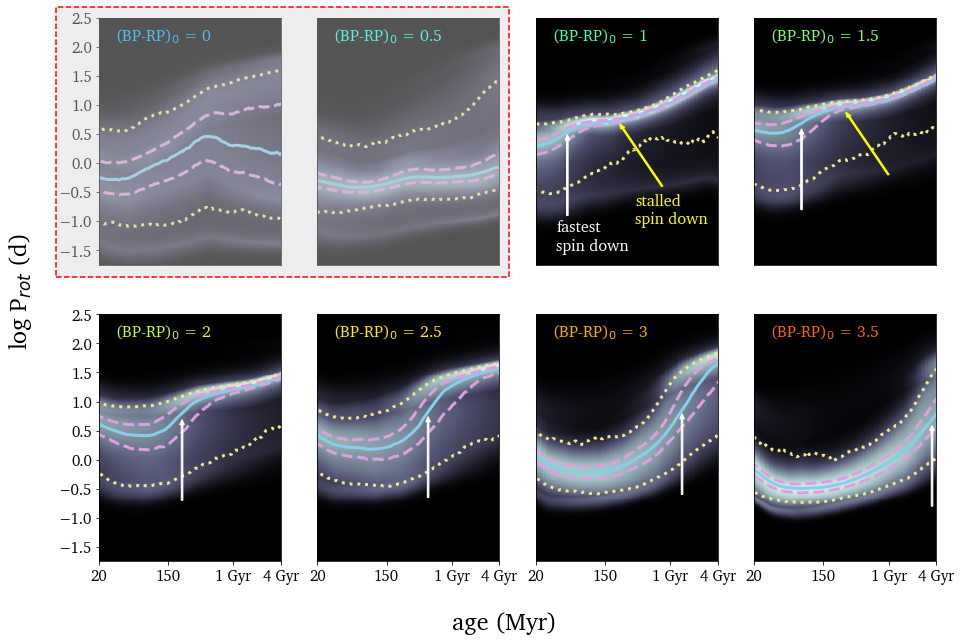}
    \caption{``Evolutionary'' tracks calculated from our model for eight different $(BP-RP)_0$ values. The background shading represents the conditional probability density $\mathcal{P}(P_{rot}\;|\;C_0,\tau)$ for a representative $\sigma_{C_0}$ value of 0.028. Contours represent the 10th, 33rd, 50th, 67th, and 90th percentiles for each panel. We have smoothed the contours for ease of interpretability by averaging across a window of 20 timesteps, which corresponds to a window width of $log (\tau)$ = 0.3. The first two panels are greyed out because of limited training data, and expected contamination from stars with radiative envelopes. In each panel, we indicate the approximate age at which the median stars experience their fastest spin down with white arrows. This age is inversely proportional to stellar mass, a ubiquitous trend in literature. We also indicate regions of apparent stalled spin down with yellow arrows.}
    \label{fig:prot_evolution_tracks}
\end{figure*}


We found an age for Theia 456 of $142^{+26}_{-21}$ Myr. \cite{Andrews...Theia456...2022AJ....163..275A} found ages for Theia 456 (using both isochrones and gyrochronology) of 150-200 Myr, and \cite{Kounkel...Covey...clusters...2019AJ....158..122K} estimated it to be $165^{+70}_{-50}$ Myr. Our estimate is on the lower end of these literature values, but consistent at $1\sigma$, so it supports the previous estimates of \citet{Andrews...Theia456...2022AJ....163..275A} and \citet{Kounkel...Covey...clusters...2019AJ....158..122K}.

We calculated NGC 6709 to be $121^{+48}_{-25}$ Myr. \cite{ColeKodikara...NGC6709...2023A&A...673A.119C} found a rotational age of $\approx$ 150 Myr for this cluster, but noted that the MS turnoff indicated an older age. They also summarized recent work as supporting ages ranging from 100 to 190 Myr. Again, our estimate is on the lower end compared to these but consistent at 1$\sigma$, so it supports the ages of \citet{ColeKodikara...NGC6709...2023A&A...673A.119C}, \citet{Lata...NGC6709...2002A&A...388..158L}, and \citet{Kharchenko...OC.ages...2005AA...438.1163K}.

Figure \ref{fig:new_cluster_densities} presents a comparison of the observed data for both systems to the probability density grids from \texttt{ChronoFlow}, analogous to Figure \ref{fig: nf_dens_old}.

\subsection{Rotational `evolution' tracks}
\label{subsec:res_rot_evol_tracks}

Another way that we can examine our data is by looking at the evolution of the $P_{rot}$ distribution at a static color. This can tell us how we would expect a group of similarly massive stars to evolve over time. Figure \ref{fig:prot_evolution_tracks} presents these results over a $(BP-RP)_0$ range of 0 to 3.5 (in steps of 0.5). It is important to note that while the contours do not necessarily represent the exact evolutionary paths that a star might take, they are a measurement of the expectation values for $P_{rot}$ over time in a population.

For $(BP-RP)_0$ = 0 and 0.5, the expected smooth spindown caused by magnetic braking is not apparent. This may be due to a lack of training data in this regime, and contamination from stars with radiative envelopes (see \S\ref{subsec:res_cluster_ages}).

The tracks at $C_0 = 1 - 2.5$ exemplify typical spindown behaviour for solar type stars. We see the initial wide distribution in rotation periods converging into a tight sequence of fast rotators, which happens slower for redder stars. Another well feature in rotation evolution is mass-dependent ``stalled spin down''. \cite{Agueros...NGC752...2018ApJ...862...33A} found evidence for this weakened braking between approximately the ages of Praesepe at 650 Myr and NGC 6811 at 1 Gyr, and \cite{Curtis...2019...rotcatalogue} and \cite{Douglas...Hyades/Praesepe...2019ApJ...879..100D} found similar results. We also see this in our model; it manifests as a flattening of the evolutionary tracks seen in Figure \ref{fig:prot_evolution_tracks} at $C_0 \approx 1-2$.

Examining the reddest stars in our sample, \cite{Popinchalk...Mdwarfrotation...2021ApJ...916...77P} found that M dwarfs at field ages exhibited a bimodal spin down pattern, with a group of fast rotators having $P_{rot} < 2$ days, and another overdensity at $P_{rot} > 60$ days. \cite{Pass...spindown...2022ApJ...936..109P} found similar results using wide binaries: some M dwarfs converged onto a slowly rotating sequence quickly, but some remained rotating fast for much longer before spinning down quickly. This bimodal spindown is also a feature of solar mass stars in physical models (e.g. \citealt{Garraffo...BimodalSpindown...2018ApJ...862...90G}). The \cite{Garraffo...BimodalSpindown...2018ApJ...862...90G} model has not been tested for fully convective M dwarfs yet, since it relies on convective turnover timescale, which has historically been difficult to measure for such stars. However, \cite{Gossage...ConvTurnTime...2024arXiv241020000G} have recently made progress towards a better method for calculating that parameter. The bimodality with fast and slow rotators is somewhat apparent in our model for $\approx$ solar mass stars, however we do not see the same buildup of fast rotators in the reddest stars. Instead, while we see a narrow distribution of slow rotators appear at $\gtrsim$ 1 Gyr, the distribution of fast rotators is broad.

From Figure \ref{fig:prot_evolution_tracks}, it is apparent that we do not expect convergence onto a slowly rotating sequence to occur over the timescale of our oldest open cluster (M67; $\approx$ 4 Gyr) for the reddest stars having $C_0 \gtrsim 3$. \cite{Pass...spindown...2022ApJ...936..109P} expect spindown for fully convective M dwarfs to occur in 2-3 Gyr. However, it is worth noting here that the delayed spindown in our model may be an artifact of the reddest stars in our sample being observed in Praesepe and Hyades (700-800 Myr), at which age they have not spun down yet. Since our catalog does not include stars having $(BP-RP)_0 \gtrsim 2.5$ past that age, it may struggle to predict their spindown. 

\subsection{Extrapolation to field ages}
\label{subsec:res_field_ages}

\begin{figure*}
    \centering
    \includegraphics[width=0.98\textwidth]{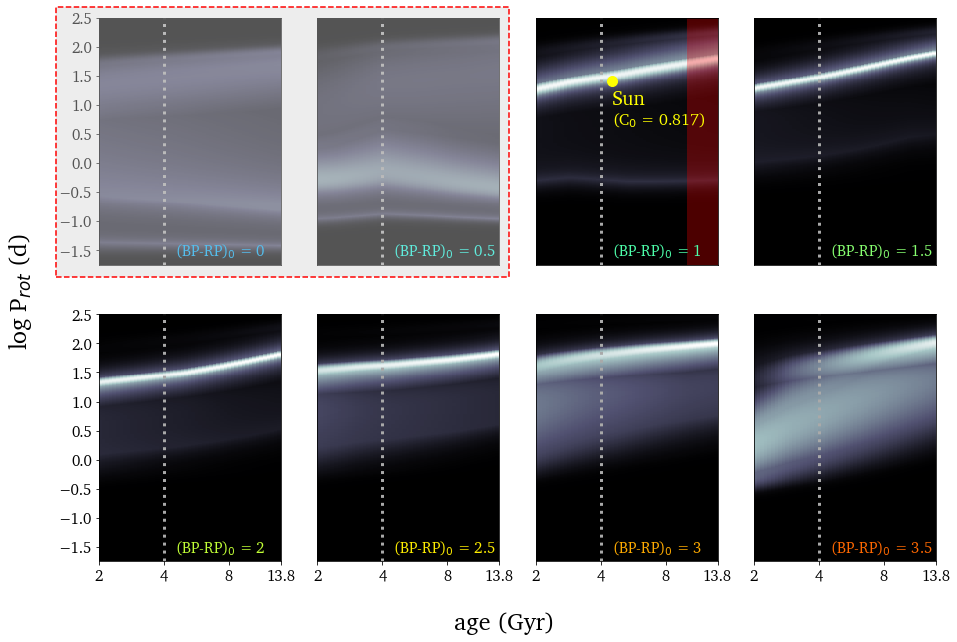}
    \caption{Probability densities $\mathcal{P}(P_{rot}\;|\;C_0,\tau)$ extrapolated to the age of the universe for different characteristic $C_0$ values. The vertical line represents are oldest training age (M67). We have plotted the Sun in the third panel for reference, showing that it fits our extrapolated model reasonably well. Note that for stars of $\approx$ similar size or larger then the Sun, they will have turned off the MS before the age of the universe, so \texttt{ChronoFlow} would not apply in that regime. This is illustrated qualitatively by the red shaded region, although the exact turnoff age would depend on stellar mass.}
     \label{fig:field_age_extrapolation}
\end{figure*}

Our oldest training stars are $\approx$ 4 Gyr. Many field stars are older than this, so the rotational evolution that \texttt{ChronoFlow} predicts past that age is extrapolation. We leave a detailed analysis for subsequent studies, but here we examine our model behavior qualitatively. Figure \ref{fig:field_age_extrapolation} presents the extrapolation of $\mathcal{P}(P_{rot}\;|\;C_0,\sigma_{C_0},\tau)$ to older ages for characteristic color values. The model evolves smoothly past 4 Gyr, and predicts a continuation of the rotational evolution seen at the ages of our older open clusters.

We do not attempt to characterize the behavior at $C_0$ = 0 or 0.5. For $C_0$ = 1 to 2, essentially all stars have converged onto a narrow slowly rotating sequence by 4 Gyr, and our model expects them to collectively continue spinning down at field ages. For $C_0$ = 2.5 to 3.5 (approximately the M dwarf regime), we see the slowest rotators continue to spin down, but also a slight narrowing of the overall distribution of $P_{rot}$. This generally aligns with the behaviour of M dwarfs at field ages characterized by \cite{Popinchalk...Mdwarfrotation...2021ApJ...916...77P} and \cite{Pass...spindown...2022ApJ...936..109P}, where some of these stars continue to rotate rapidly at field ages.

As a simple test of our model's extrapolation behavior, as described in \S\ref{subsec:res_single_star_ages}, we obtained an age estimate for the Sun of $5.1^{+1.7}_{-1.4}$ Gyr.

\section{Conclusions and Summary} \label{sec: conclusions}

In this work we constructed a rotational catalog of 7,615 stars, from 28 open clusters plus subgroups of the Taurus and Sco-Cen associations. This is the most comprehensive open cluster rotational catalog to date, spanning an age range of 1.5 Myr to 4 Gyr. We have standardized all photometry to Gaia DR3, which we de-reddened using the 3D dustmaps from \cite{Edenhofer2023...dustmap...2023arXiv230801295E} and \cite{Bayestar19...dustmap...2019ApJ...887...93G}. We also propagated uncertainties in parallax, Gaia photometry, and dustmap extinction to get realistic color uncertainties. Finally, we have included cluster membership probabilities, sourced from literature and HDBScan analysis. We present this catalog as part of this work, as well as a deep literature compilation of age estimates for the clusters in the catalog. We hope that this catalog will be a useful benchmark for gyrochronology, and also for the broader astronomical community.

We used this catalog to develop a completely data-driven gyrochronology model using a conditional normalizing flow: \texttt{ChronoFlow}. We trained \texttt{ChronoFlow} to learn $P_{rot}$ probability density distributions conditioned on Gaia DR3 color, photometric uncertainty, and age. By constructing it in this way, \textit{we insulate the model against observational biases in color and age space.} Our model is also the first to use variable cluster membership probabilities per star, which constrains our model more realistically.

We have shown that \texttt{ChronoFlow} can model the $P_{rot}$ dispersion observed in similar coeval stars across a larger parameter space at open cluster ages than existing literature models (although it does not extend into field ages as \texttt{GPgyro} does). Furthermore, it is at least as accurate as existing models at recovering stellar ages without introducing systematic bias due to model structure. Our model can be used to compute the probability density in color-$P_{rot}$ space at any given age, or examine the $P_{rot}$ density evolution over time for fixed colors. We have also implemented \texttt{ChronoFlow} in a Bayesian framework to \textit{infer stellar and cluster age posterior probabilities}. The uncertainty on individual stellar ages is $\approx$ 0.7 dex, and we can constrain cluster ages with a statistical uncertainty of $\approx$ 0.06 dex, or 15\%. As an example of its applicability, we have used \texttt{ChronoFlow} to calculate new age estimates for five open clusters and one stellar stream.

Furthermore, we thoroughly tested systematic uncertainties. These includes the effect of mixing dustmaps when calculating stellar extinction, the effect of cluster member selection, a comparison of space vs. ground based surveys, and the variance caused by different families of training ages. We note two dominant sources of systematic uncertainty: (i) the application of different extinction procedures to photometry for training vs. inference, and (ii): the choice of ages in the training dataset. Both of these affect cluster age inference to the order of $\approx$ 0.04 dex, and the total uncertainty arising from all sources of systematics is $\approx$ 0.06 dex. Combined with the statistical uncertainty of our model, our total uncertainty in population age inference \textbf{$\approx$ 0.08 dex}.

To summarize the key outcomes of this work:

\begin{itemize}
    \item We have developed the most robust open cluster rotational catalog to date, with \textbf{standardized Gaia DR3 photometry} de-reddened using 3D dust maps, and \textbf{accurate photometric errors}.
    \item We provide \textbf{cluster membership probabilities} for all stars in our catalog.
    \item We have developed \textbf{\texttt{ChronoFlow}}: a data-driven model that accurately predicts rotational evolution probabilistically.
    \item \texttt{ChronoFlow} incorporates cluster membership probabilities per star, and \textbf{is robust against observational selection effects in color and age.}
    \item With \texttt{ChronoFlow}, we can estimate stellar ages to a statistical uncertainty of \textbf{0.7 dex}, and cluster ages to \textbf{0.06 dex}.
    \item We have analyzed sources of possible systematic error, which in total add \textbf{0.06 dex} of uncertainty to inferred population ages. This results in a total error budget of \textbf{0.08 dex}.
\end{itemize}

\noindent \texttt{ChronoFlow} can be broadly applied as an age inference tool with confidence in the consideration of systematic uncertainties. It also represents a baseline that can now be used to probe and compare theoretical models, providing insight into the physics that drive stellar spin down.


\section{Acknowledgments} \label{sec:ack}

P.R.V. was supported by funding from the Dunlap Institute, a Queen Elizabeth II Graduate Scholarship in Science and Technology from the Government of Ontario, and a Canada Graduate Scholarships – Master’s (CGS-M) Award and Postgraduate Scholarships - Doctoral (PGS-D) Award from the Natural Sciences and Engineering Research Council of Canada (NSERC). 
J.S.S. was supported by funding from the Dunlap Institute, an NSERC Banting Postdoctoral Fellowship, NSERC Discovery Grant RGPIN-2023-04849, and a University of Toronto Connaught New Researcher Award. 
This work was supported in part by a NSERC Discovery Grant (RGPIN-2020-04554) to G.M.E.

The authors would like to thank Diego Godoy-Rivera for inspiration and helpful feedback on this work. The authors would also like to thank Cecilia Garraffo and Kristina Monsch for hosting P.R.V. and J.S.S. at CfA for valuable collaboration time, and their feedback and ideas for this project. The authors would like to thank Steffani Grondin for valuable feedback on this paper, and Ryan Cloutier, Jennifer van Saders, Dario Fritzewski, Jason Curtis, and David Charbonneau for useful comments and insights.

\bibliography{refs}
\bibliographystyle{aasjournal}


\appendix

\section{Cluster CMDs and proper motion plots}
\label{app:cluster_summary_plots}

Figure \ref{fig: CMDs} presents CMDs for each cluster, comparing observations to the theoretical isochrones generated using \texttt{brutus}\footnote{\url{https://github.com/joshspeagle/brutus/tree/master}} (which implements MIST isochrones; \citealt{MIST...2016ApJS..222....8D}) at the fiducial literature age for each cluster. The red triangles indicate the stars we exclude from our catalog due to the final CMD cut. Figure \ref{fig: proper_motions} shows the distributions in proper motion of catalog stars compared to field stars around each cluster.

\begin{figure*}[!hbt]
    \centering
    \includegraphics[width=0.94\textwidth]{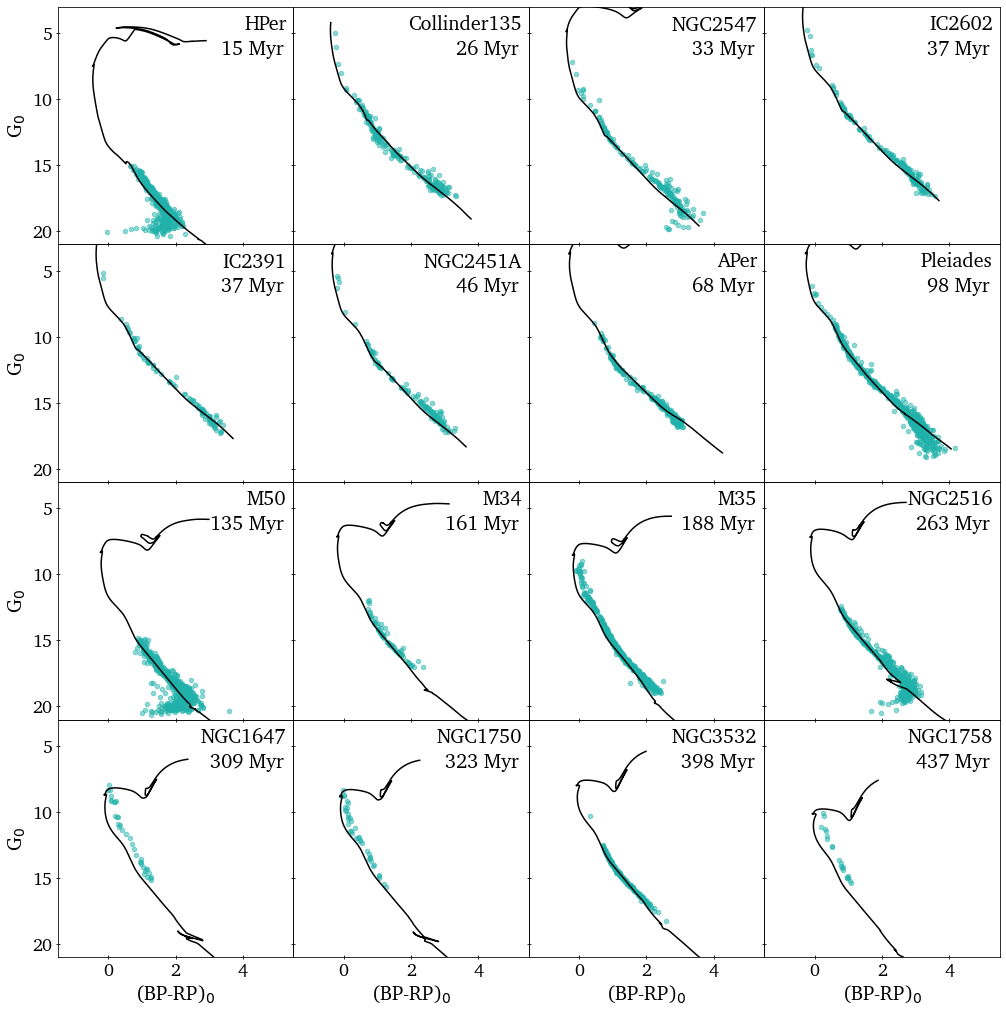}
\end{figure*}

\begin{figure*}
    \ContinuedFloat
    \centering
    \includegraphics[width=0.94\textwidth]{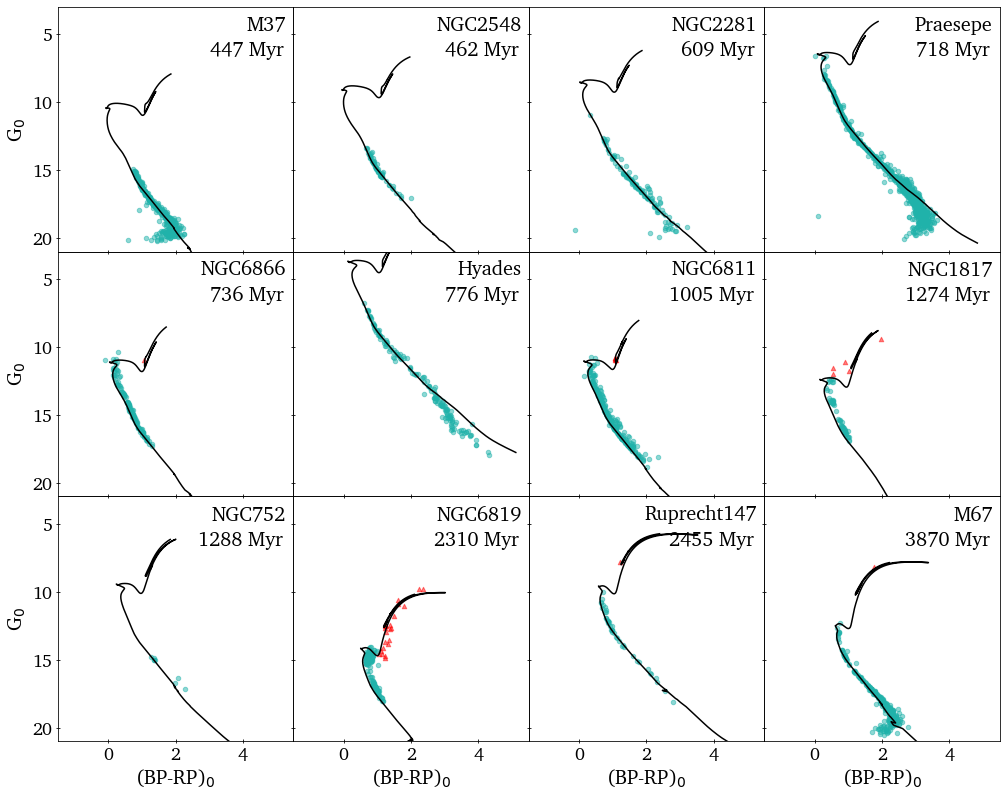}
    \caption{Cluster CMDs in increasing order of cluster age (excluding the Taurus and Sco-Cen regions). Both the $G_0$ magnitudes and $(BP-RP)_0$ colours are de-reddened using the extinctions from \cite{Edenhofer2023...dustmap...2023arXiv230801295E}. Isochrones are generated using the fiducial cluster ages from literature; they are not estimates inferred by our model. Red triangles are stars that we have excluded from our catalog due to their location on the CMDs.}
    \label{fig: CMDs}
\end{figure*}    

\clearpage

\begin{figure*}[!hbt]

    \centering
    \includegraphics[width=0.98\textwidth]{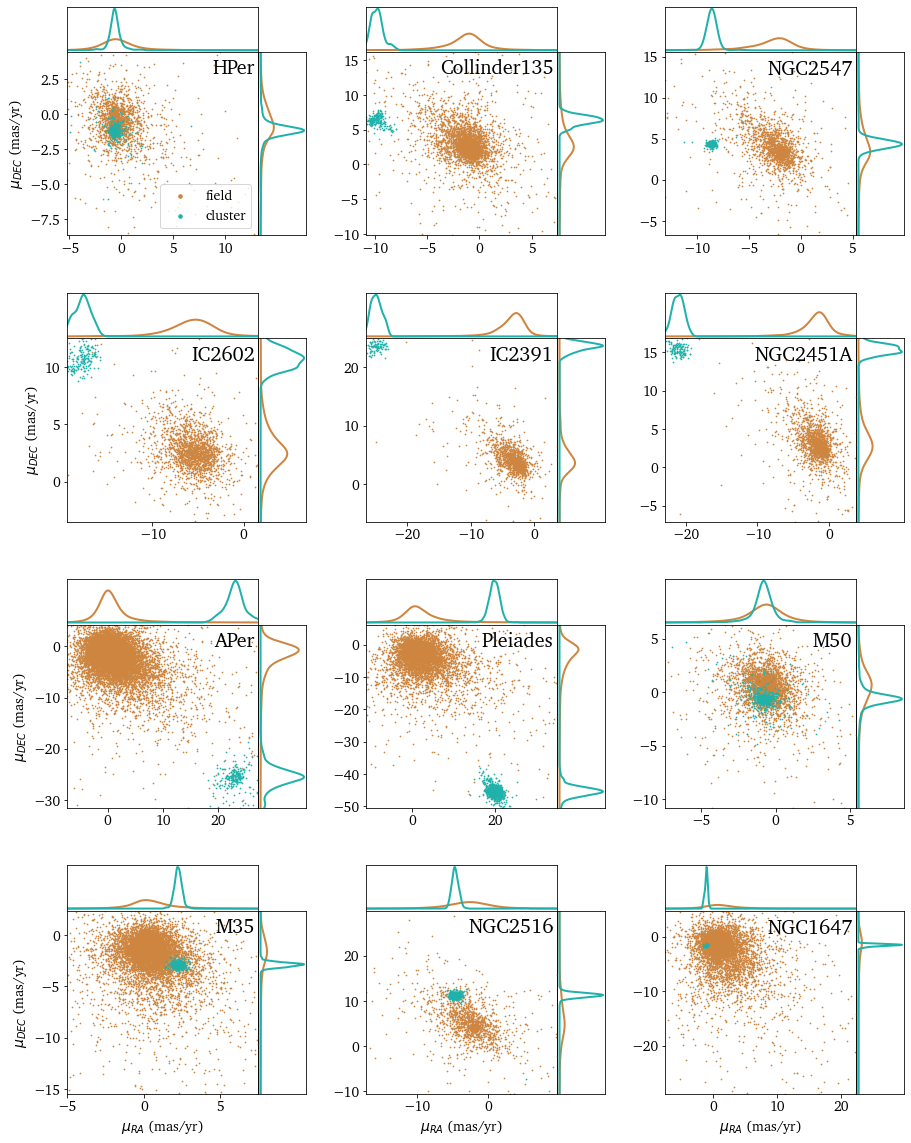}

\end{figure*}

\begin{figure*}
    \ContinuedFloat
    \centering
    \includegraphics[width=0.98\textwidth]{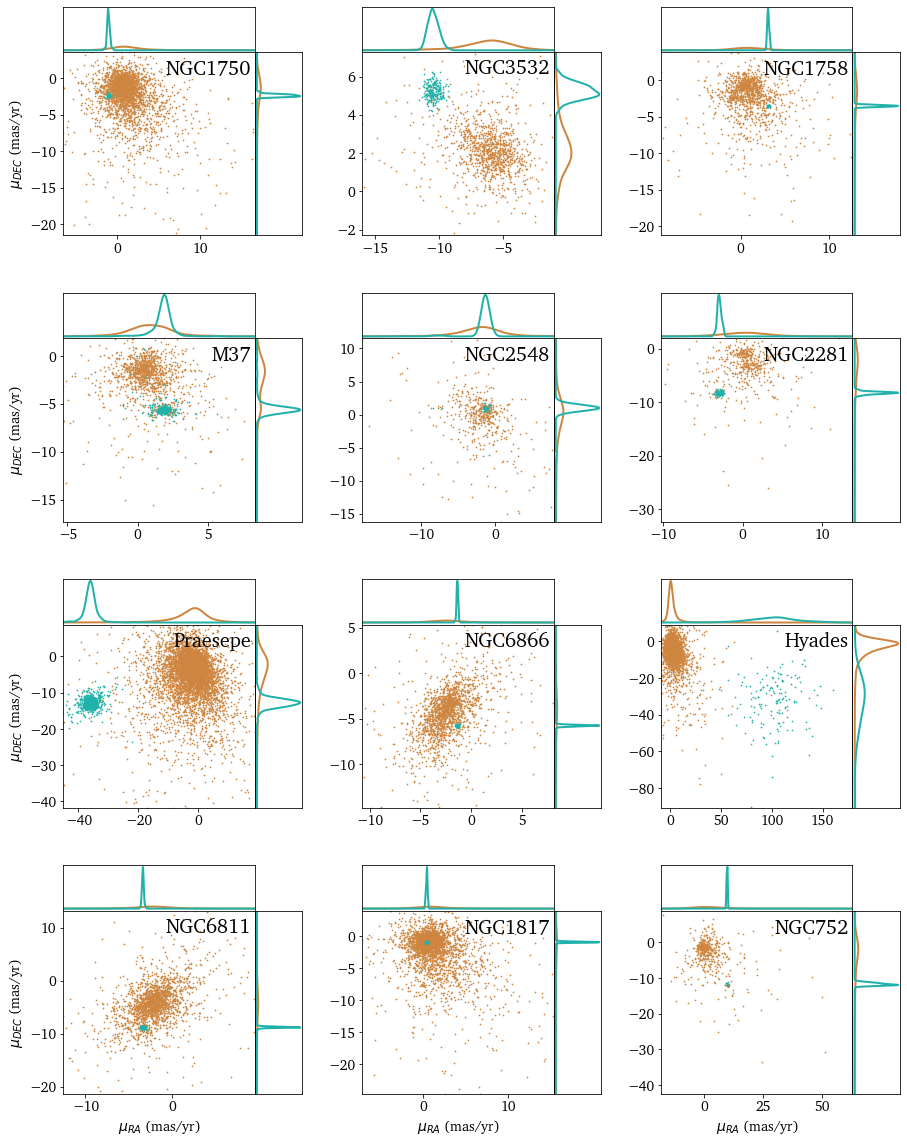}
\end{figure*}

\begin{figure*}[!ht]
    \ContinuedFloat
    \centering
    \includegraphics[width=0.98\textwidth]{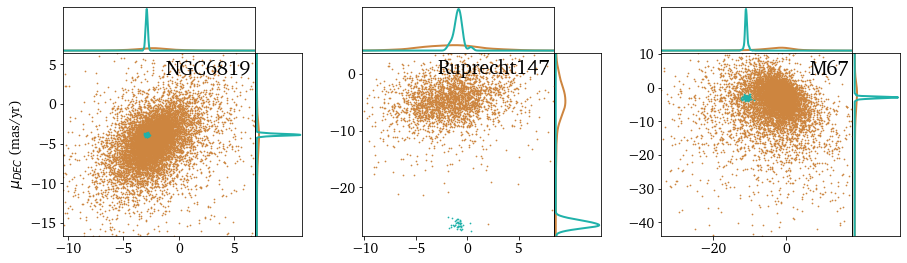}
    \caption{Proper motion distributions for each cluster. Cluster members are plotted against a 2000-star sample drawn randomly from between 4$\sigma$ and 10$\sigma$ in distance from the cluster center. Marginal distributions for $\mu_{RA}$ and $\mu_{DEC}$ are plotted along the axes for both cluster members (teal) and field stars (brown). The young subgroups of the Taurus and Sco-Cen associations are excluded from this plot.}
    \label{fig: proper_motions}
\end{figure*}

\section{Cuts applied to literature rotation catalogs}
\label{app:vetting_all}

Here we summarize quality cuts we made when compiling our catalog, based only on the information provided by the literature sources. Many of these are designed to eliminate binaries.

\begin{itemize}
    \item \cite{Curtis2020_gyro} already applied quality cuts when compiling their rotators, so we use that catalog (Table 5 in that work) with no additional filters for NGC 752, NGC 6811, NGC 6819, Pleiades, and Ruprecht 147. We do not use the Praesepe data from this catalog, and instead use \cite{Rampalli...Praesepe...2021ApJ...921..167R}, which is a revised catalog built on the \citet{Curtis2020_gyro} data.
    \item We do not apply any additional cuts to the \cite{Barnes...gyro...NGC2548...2015AA...583A..73B} NGC 2548 catalog (Table 2 in that work).
    \item We do not apply additional cuts to the \cite{Fritzewski...gyro...NGC2281...2023AA...674A..152F} NGC 2281 catalog (Table 2 in that work).
    \item We exclude stars in the \cite{Douglas...Hyades/Praesepe...2019ApJ...879..100D} Hyades catalog (Tables 2/3 of that work) having a \texttt{Conf} flag of \texttt{Y} (ie. confirmed binaries).
    \item We exclude all rotators from the \cite{GodoyRivera...2021...rotcatalogue} catalog (Table F1 of that paper) that were assigned the classification of \texttt{non-member}.
    \item We only include rotators from the \cite{Douglas...SouthClusters...2024ApJ...962...16D} catalog (Table 4 of that work) that have a \texttt{to\_plot} value of 1; this was used as a membership cut in that work.
    \item We apply several cuts to the \cite{Long_gyrocatalogue...2023ApJS..268...30L} catalog (Table 5 of that work):
        \begin{enumerate}
            \item We only include stars with \texttt{BinPhot} and \texttt{BinSpec} values that are both 0 (ie. photometrically and spectroscopically single stars).
            \item We only include stars with a \texttt{BinRUWE} value of 0. This corresponds to stars with a Gaia \texttt{RUWE} value of $<$1.4, which is a cut we apply in our final selection anyways.
            \item We only include stars with a \texttt{BinFlag} value of 0 or 1 (either a MS single star or unknown).
            \item We also exclude four Pleiades stars that \texttt{L23} sourced directly from the \cite{Hartman...Pleiades...2010MNRAS.408..475H} catalog, but for which the original rotation period was measured in other works (\citealt{Prosser...Pleiades...1993PASP..105.1407P}; a reference labelled P95 which we assume to be \citealt{Prosser...Pleiades...1995PASP..107..211P}; and \citealt{Messina...Pleiades...2001AA...371.1024M}), and for which \cite{Hartman...Pleiades...2010MNRAS.408..475H} were not able to measure rotation periods themselves. 
        \end{enumerate}
        \item We only include rotators from \cite{BoyleBouma...gyro.Aper...2022arXiv221109822B} (Table 3 in that work) where the \texttt{in\_gyro\_sample} flag is set to \texttt{True}. These are the subset of stars that the authors use to calibrate their model.
        \item We only keep stars from \cite{Moraux...gyro...HPer...2013AA...560A..13M} (Table 2 in that work) with a \texttt{bin} flag of 1. This corresponds to photometrically single stars.
        \item We only keep stars from \cite{Meibom...M34...2011ApJ...733..115M} (Table 2 in that work) that are kinematic and/or photometric members: i.e. having a \texttt{mcode} value of \texttt{KM} or \texttt{PM}.
        \item We only include stars with a \texttt{single} flag of \texttt{True} and a \texttt{binary} flag of \texttt{False} from \cite{Dungee...gyro.M67...2022ApJ...938..118D} (Table 2 in that work). We also exclude stars with a false-alarm probablity (FAP) of $\geq$ 1\%, which the authors use as the upper limit for which periodic signals are considered significant. We exclude stars with a measured period of $\geq$ 175 days, following the authors' requirement that stars have at least five completed periods within the light-curve duration to be considered reliable.
        \item We only include stars from \cite{Barnes...M67...2016ApJ...823...16B} (Table 1 in that work) having a \texttt{Member} flag of \texttt{SM} (i.e. single members).
        \item We only include stars from \cite{Rebull...gyro.USco.ROph...2018AJ....155..196R} (Table 1 of that work) having a \texttt{Rdist} flag of \texttt{no}, meaning that there were not resolved distant peaks in the periodogram, which the authors would interpret as a binary. Although resolved close peaks could also be indicative of binaries, they could result from differential surface rotation, so we do not apply a cut based on the latter phenomenon.
        \item We only include stars from \cite{Rampalli...Praesepe...2021ApJ...921..167R} (Table 3 of that work) having a \texttt{Bin} flag of 0 (i.e. non-binary stars) and a \texttt{QFClean} flag of 1. The latter cut restricts our sample to only stars with high-quality light curve flags.
        \item We only include stars from \cite{Fritzewski...gyro.NGC3532...2021AA...656A.103F} catalog (Table 2 of that work) having a \texttt{Binary} flag of 0.
        \item We exclude stars from the \citet{Fritzewski...NGC2516...2020AA...641A..51F} catalog (NGC 2516) that are included in Table 2 of that work as potential radial velocity binaries. 
\end{itemize}

\section{Notes and edge cases from rotation period compilation} \label{sec:app_prot_notes}

Here we describe edge cases and duplication that we encountered when compiling our rotator catalog.

\subsection{Multiple catalogs using the same source measurement}

Here is an extended description of the rotation periods that were cited by multiple source catalogs (as described in \S\ref{sec: data}:

\begin{itemize}
    \item Both \cite{Long_gyrocatalogue...2023ApJS..268...30L} and \cite{Douglas...Hyades/Praesepe...2019ApJ...879..100D} used rotation periods measured by both \cite{Delorme...Hyades...Praesepe...2011MNRAS.413.2218D} and \cite{Hartman...HATNet...2011AJ....141..166H} in their Hyades rotator catalogs.
    \item \cite{Long_gyrocatalogue...2023ApJS..268...30L} cited M67 rotation periods directly from \cite{Dungee...gyro.M67...2022ApJ...938..118D}, however we consider these to be different catalogs since \cite{Long_gyrocatalogue...2023ApJS..268...30L} calculated their own set of M67 rotation periods as well.
    \item Both \citet{GodoyRivera...2021...rotcatalogue} and \citet{Fritzewski...NGC2516...2020AA...641A..51F} used rotation periods measured by \citet{Irwin...NGC2516...2007MNRAS.377..741I} for NGC 2516.
    \item Similarly, \cite{Long_gyrocatalogue...2023ApJS..268...30L} cited Ruprecht 147 rotation periods directly from \cite{Curtis2020_gyro}, along with measuring their own rotation periods.
    \item \cite{GodoyRivera...2021...rotcatalogue} and \cite{Curtis2020_gyro} both cited NGC 6811 rotation periods directly from \cite{Curtis...2019...rotcatalogue}.
    \item \cite{GodoyRivera...2021...rotcatalogue} and \cite{Curtis2020_gyro} both cited Pleiades rotation periods directly from \cite{Rebull...Pleiades...2016AJ....152..113R}.
\end{itemize}

\subsection{Hyades}
\label{subsec:Prot_Hyades}

Both the \cite{Douglas...Hyades/Praesepe...2019ApJ...879..100D} and \cite{Long_gyrocatalogue...2023ApJS..268...30L} catalogs use rotation periods measured by \cite{Hartman...HATNet...2011AJ....141..166H}. \cite{Long_gyrocatalogue...2023ApJS..268...30L} include one rotator and \cite{Douglas...Hyades/Praesepe...2019ApJ...879..100D} include four, (of which one is the rotator used by \citealt{Long_gyrocatalogue...2023ApJS..268...30L}). Both catalogs also share eight rotators measured by \cite{Delorme...Hyades...Praesepe...2011MNRAS.413.2218D}, of which two have a disagreement in reported period as cited in the catalogs. In both cases, the values reported by \cite{Douglas...Hyades/Praesepe...2019ApJ...879..100D} match those reported directly by \cite{Delorme...Hyades...Praesepe...2011MNRAS.413.2218D}, so we use these values. \cite{Long_gyrocatalogue...2023ApJS..268...30L} also include one rotator that is not part of the \cite{Douglas...Hyades/Praesepe...2019ApJ...879..100D} catalog; \cite{Douglas...Hyades/Praesepe...2019ApJ...879..100D} include 16 that are not part of the \cite{Long_gyrocatalogue...2023ApJS..268...30L} catalog.

\subsection{M35}
\label{subsec:Prot_M35}

\cite{Long_gyrocatalogue...2023ApJS..268...30L} reported rotation periods for M35 from a number of sources which overlap. In particular, they cite rotation periods directly from \cite{Meibom...M35...2009ApJ...695..679M}, \cite{Libralato...M35...2016MNRAS.456.1137L}, and \cite{Jeffries...M35...2021MNRAS.500.1158J}. Since \cite{Jeffries...M35...2021MNRAS.500.1158J} themselves compiled their rotation periods from both \cite{Meibom...M35...2009ApJ...695..679M} and \cite{Libralato...M35...2016MNRAS.456.1137L}, we checked for duplicates. We found none, so no further analysis was required.

\subsection{M67}
\label{subsubsec:Prot_M67}

\cite{Long_gyrocatalogue...2023ApJS..268...30L} used 10 stars with rotation periods measured by \cite{Dungee...gyro.M67...2022ApJ...938..118D} that were not considered in the final \cite{Dungee...gyro.M67...2022ApJ...938..118D} catalog. One of these has a measured rotation period but a FAP $>$ 0.01 (which was used as a quality cut by \citealt{Dungee...gyro.M67...2022ApJ...938..118D}). The other nine stars do not actually have rotation periods reported by \cite{Dungee...gyro.M67...2022ApJ...938..118D} catalog, so we were not able to determine where \cite{Long_gyrocatalogue...2023ApJS..268...30L} obtained these values from from. As such, we exclude these 10 stars from our final catalog. The other 45 rotation periods cited by \cite{Long_gyrocatalogue...2023ApJS..268...30L} from \cite{Dungee...gyro.M67...2022ApJ...938..118D} agree with the values presented by the latter.

\subsection{NGC2516}
\label{subsec:Prot_NGC2516}

There appears to be duplication in the NGC 2516 data from \cite{Irwin...NGC2516...2007MNRAS.377..741I} as compiled by \cite{GodoyRivera...2021...rotcatalogue}. Stars \texttt{N2516-2-1-1441} and \texttt{N2516-1-8-1419} are crossmatched to the same Gaia DR2 ID, and have similar rotation periods (1.461 and 1.465 days, respectively). We therefore remove the latter entry from our catalog. This also occurs with \texttt{N2516-1-8-862} and \texttt{N2516-2-1-843} (rotation periods of 1.545 and 1,546 respectively), so we removed the former from our catalog. We saw this again with \texttt{N2516-1-8-1300} and \texttt{N2516-2-1-1317} (both having a rotation period of 0.37 days), and removed the former. Finally, we saw this with \texttt{N2516-1-8-2732} and \texttt{N2516-2-1-2854} too (both having rotation periods of 0.341 days), and removed the former.

\subsection{NGC6811}
\label{subsec:Prot_NGC6811}

\cite{GodoyRivera...2021...rotcatalogue} include 34 rotators measured by \cite{Curtis...2019...rotcatalogue}, however 10 of their reported periods differ from the C19 values. In these cases we use the rotation periods directly from \citet{Curtis...2019...rotcatalogue}. Additionally, they include five rotators from the \cite{AS+21_Kepler_catalog,AS+19_Kepler_catalog} catalogs that were included in preliminary versions but not in the final published versions of those catalogs for various reasons. From private communication with Diego Godoy-Rivera, we determined that KIC 9656397 was a F-type eclipsing binary which should not be included in our sample, but the other four were reliable measurements which we kept.

\subsection{Ruprecht 147}
\label{subsec:Prot_Ruprecht147}

The \cite{Long_gyrocatalogue...2023ApJS..268...30L} catalog directly uses the rotation period measurements for 12 stars from \cite{Curtis2020_gyro}.
Out of these, we found seven matches in \cite{Curtis2020_gyro} catalog based on our crossmatch from \cite{Long_gyrocatalogue...2023ApJS..268...30L} Gaia DR3 IDs to \cite{Curtis2020_gyro} Gaia DR2 IDs. For all seven, the rotation period is accurately reported. For the other five stars, no match was found in \cite{Curtis2020_gyro} catalog and none of the periods reported in \cite{Long_gyrocatalogue...2023ApJS..268...30L} are close matches to periods in \cite{Curtis2020_gyro} catalog, so we exclude these rotation periods from our catalog.

\section{Notes and edge cases from Gaia DR3 crossmatch}
\label{app:sec_dr3_crossmatch}

Here we describe the steps we took to crossmatch all rotation periods to Gaia DR3:

\begin{enumerate}
    \item If a star had a DR2 ID in the source catalog, we found the nearest neighbors in DR3 using the \texttt{gaiaedr3.dr2\_neighbourhood} table. If there are multiple matches, we consider the angular and magnitude differences, and select the best match based on both criteria. We find a match for 2,495 rotators this way.

    \item If the star had a DR2 ID but no neighbor in the above step, or did not have a Gaia DR2 ID in the source catalog, we used \texttt{topcat} \citep{Topcat...2005ASPC..347...29T} and/or the Gaia archive search to find the closest match within 1" using the R.A. and Decl. values provided with the catalog. In some cases this radius had to be extended to 2" to find a match.

    \item There were 89 stars in the \cite{GodoyRivera...2021...rotcatalogue} catalog for which no R.A. or Decl. was provided in that paper, but for which we found measurements in the rotation source catalogs for those rotators. 17 of these had Gaia DR3 matches within 1", however four of those 17 were ambiguous so we exclude those from our sample. The other 72 had matches within 10", however we only include the 25 of those which were unambiguous in our final catalog.

    \item For H Persei, 483 of the 507 stars in the \cite{Moraux...gyro...HPer...2013AA...560A..13M} catalog had Gaia DR3 crossmatches within 1as. An additional 6 had crossmatches within 5as, and 15 had crossmatches between 5 and 10as. At these larger radii, since the stars often had multiple crossmatches, we excluded these from our catalog and retain the 483 initial 1" crossmatches.

    \item The crossmatches were all double checked using apparent magnitudes in the G and V bands where available (we converted G to V and vice versa where necessary using the $G-V$ and $B-V$ relationships presented in table 5.8 of \citealt{GaiaDR2...documentation...2018gdr2.reptE....V}). In some cases, better matches were found based on a slightly larger angular difference but smaller difference in apparent magnitude. 15 stars were manually corrected to a different Gaia DR3 after this magnitude comparison.

\end{enumerate}

There were 172 stars that corresponded to members of different clusters in different catalogs (but matched to the same Gaia DR3 ID). 169 of these were classified as Upper Scorpius by \cite{Rebull...gyro.USco.ROph...2018AJ....155..196R} catalog but $\rho$ Ophiucus by \cite{Long_gyrocatalogue...2023ApJS..268...30L}. The rotation period measurements were very consistent for each. This is not an issue since we assigned all members of the Sco-Cen association to subgroups using the \cite{Ratzenbock...ScoCen_groups...2023AA...677A..59R} classifications anyway. In two other cases, one record was a direct match based on the Gaia DR3 ID, and one was a best fit match based on the DR2 ID. In these cases we kept the record with the matching DR3 ID. In the last case, the same DR3 ID was present in the source catalog for two different clusters. We excluded this star from all further analysis.

We also exclude two stars from our catalog which are members of Taurus subgroups according to \cite{Krolikowski...Taurus...2021AJ....162..110K}, but are part of the \cite{Douglas...Hyades/Praesepe...2019ApJ...879..100D} and \cite{GodoyRivera...2021...rotcatalogue} Hyades and Pleiades catalogs respectively.

\section{Cluster membership probabilities}
\label{app:cluster_membership_probs_details}

Here we describe how we assigned cluster membership probabilities to rotators where no quantitative membership probability was provided in the data source catalog.

\subsection{Godoy Rivera et al. (2021): M37, M50, NGC 2516, NGC 2547, NGC 6811, Pleiades, Praesepe}

The \cite{GodoyRivera...2021...rotcatalogue} catalog contains cluster membership probabilities for each star, except those with insufficient Gaia DR2 data to be assigned a membership probability. These stars, designated as part of the \textit{no-info} group by \cite{GodoyRivera...2021...rotcatalogue}, we assign a probability of 0.5 to.

\subsection{Fritzewski et al. (2020): NGC 2516}

\citet{Fritzewski...NGC2516...2020AA...641A..51F} considered four membership criteria (photometry, proper motion, parallax, and radial velocity), and only included stars in their rotator catalog that passed at least two of those in their rotator catalog. Since they did not estimate quantitative membership probabilities per star, we therefore assign a default value of 0.9 for all of those stars.

\subsection{Douglas et al. (2024): Collinder 135, IC 2391, IC 2602, NGC 2451A, NGC 2547}

For the \cite{Douglas...SouthClusters...2024ApJ...962...16D} sources, we adopt the HDBSCan membership probability provided for each star as the cluster membership probability. The exception to this is Gaia DR3 ID 5602911354490249344, which has no HDBScan membership probability provided or GES membership probability, but a \cite{CantatGaudin...catalogue...2020AA...640A...1C} value of 1, which we adopt.

\subsection{Douglas et al. (2019): Hyades and Praesepe}

\cite{Douglas...Hyades/Praesepe...2019ApJ...879..100D} use a membership catalog compiled from several sources. Since no individual stellar cluster membership probabilities are provided there, we use a default value of 0.9. It is worth notating that some of these stars were not classified as cluster members according to \cite{GaiaCollab...2018AA...616A..10G}, however this may be due to Gaia quality cuts, and other good measurements are available.

\subsection{Rampalli et al. (2021): Praesepe}

\cite{Rampalli...Praesepe...2021ApJ...921..167R} also compile a membership catalog from multiple sources, and define all as having a $>$70\% cluster membership probability. Naively assuming a uniform distribution in probability space, we assign a value of 0.85 for all of these stars.

\subsection{NGC 3532}

For NGC 3532, we use the radial velocity probabilities as calculated by  \cite{Fritzewski...NGC3532specmembership...2019AA...622A.110F} as our cluster membership probabilities. Where these are not available (this is the case for 9 of the 139 stars ), we adopt a probability of 0.9.

\subsection{NGC 2281}

No quantitative membership probabilities were provided by \cite{Fritzewski...gyro...NGC2281...2023AA...674A..152F} for NGC 2281, however they considered photometry, astrometry and radial velocities in their expansion of the NGC 2281 catalog from \cite{Kounkel...Covey...clusters...2019AJ....158..122K}, which also did not provide quantitative membership probabilities. We adopt 0.9 as the default value for this cluster.

\subsection{Barnes et al. (2016): M67}

\cite{Barnes...M67...2016ApJ...823...16B} use the M67 membership catalog from \cite{Geller...M67...2015AJ....150...97G}, which computers radial velocity cluster membership values, and compares proper motion cluster membership probabilities from four literature sources. We take the average value across the four sources as the proper motion cluster membership probability, and average that with the RV membership probability for each star to get our final value. Two stars don't have any proper motion probabilities from literature; for these we assign the RV membership probability.

\subsection{NGC 2548}

\cite{Barnes...gyro...NGC2548...2015AA...583A..73B} use photometry to determine cluster membership probabilities. They also provide the cluster member classification from \cite{BalaguerNunez...NGC2548...membership...2005AA...437..457B}, however not every star has a crossmatch in that catalog. Therefore we use the classification of "provisional members" and "confirmed members", to which we assign membership probabilities of 0.7 and 0.9 respectively. 

\subsection{M34}

In the \cite{Meibom...M34...2011ApJ...733..115M} M34 catalog, stars are either classified as kinematic (and photometric) members, or just photometric members. For the kinematic members, a radial velocity membership probability is provided, and in some cases a photometric probability is too. We take the average of these where both are available, and take the radial velocity value $P_{mem,RV}$ where not. We assign a membership probability of 0.7 to the only-photometric members.

\subsection{H Persei}

\cite{Moraux...gyro...HPer...2013AA...560A..13M} compiles membership data from \cite{Currie...HPer...2010ApJS..186..191C}, who analyze both spectroscopic and photometric membership. \cite{Currie...HPer...2010ApJS..186..191C} estimate a 16\% contamination rate in their photometric catalog, so we assign a value of 0.84 cluster membership probability for their (only) photometric members. We also assign a membership probability of 0.84 for spectroscopic only members, and 0.9 for stars that are both photometric and spectroscopic members. For stars in the \cite{Moraux...gyro...HPer...2013AA...560A..13M} catalog that they do not report a membership category from \cite{Currie...HPer...2010ApJS..186..191C} from, we assign a value of 0.7, as no quantitative membership probabilities are provided by \cite{Moraux...gyro...HPer...2013AA...560A..13M}. For non-members according to \cite{Currie...HPer...2010ApJS..186..191C}, we assign a value of 0.5.

\subsection{Curtis et al. (2019): NGC 6811}

For NGC 6811, \cite{Curtis...2019...rotcatalogue} applu astrometric cuts, followed by photometric cuts applied on a Gaia G vs. BP-RP CMD. They do not provide individual cluster membership probabilities, so we assign a value of 0.9 to all stars.

\subsection{Curtis et al. (2020): NGC 752, NGC 6819, Pleiades, Ruprecht 147}

For the Pleiades, NGC 752, and NGC 6819, \cite{Curtis2020_gyro} applied Gaia DR2 astrometry, RVs and CMDs to identify single star members in the same method as for Ruprecht 147. For Ruprecht 147, they used proper motion candidates from literature, filtered that list using radial velocities, and merged further Gaia DR2-based catalogs from literature. We assign a cluster membership probability of 0.9 to all of these candidates.

\subsection{Upper Scorpius and $\rho$ Ophiucus subgroups}

As part of their census of the Sco-Cen region, \cite{Ratzenbock...ScoCen_groups...2023AA...677A..59R} provide ``stability values'' for the membership of each Gaia DR3 star in their corresponding groups. We use this as our cluster membership probability.

\subsection{Taurus}

\cite{Krolikowski...Taurus...2021AJ....162..110K} do not provide cluster membership probabilities for the Taurus sub-groups, so we adopt a value of 0.9. For most of these groups, the ages are very similar so the effect of miscategorizing them would be negligible.

\section{Model Performance Tests}
\label{app:model_tests}

Here we provide further details on the tests we executed to evaluate the performance of \texttt{ChronoFlow}, as described briefly in \S\ref{sec: results}. We used the average stellar log probability as a benchmark metric to \textbf{measure} how well observations aligned with our $P_{rot}$ density probabilities (e.g., Figures \ref{fig: nf_dens_old} and \ref{fig: nf_dens_young}). Averaging the stellar log probabilities \textbf{accounts} for the different sample sizes of each cluster.

\subsection{Model performance by cluster} \label{subsec:res_violin_cluster_probs}

\begin{figure}
    \centering
    \includegraphics[width=0.82\textwidth]{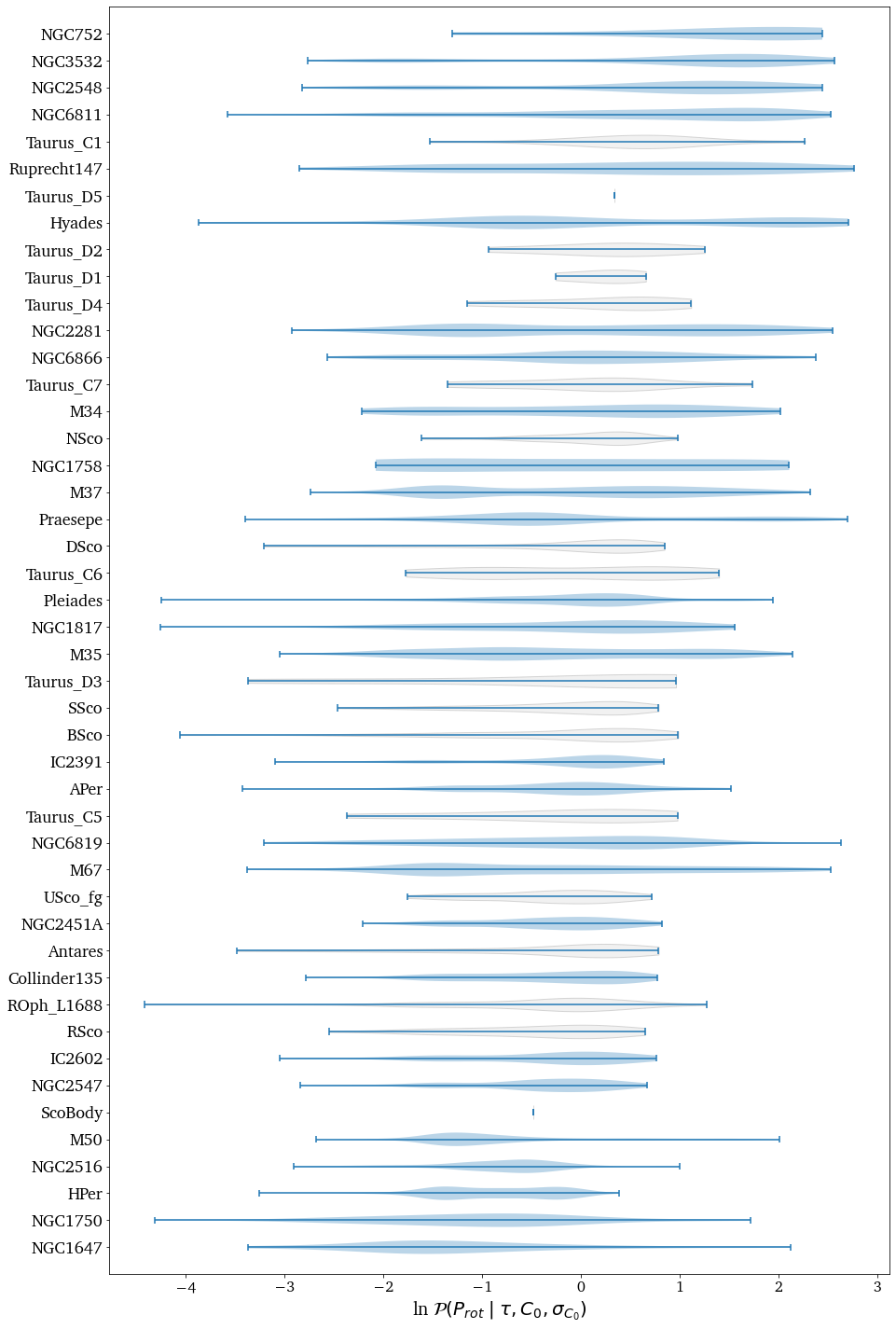}
    \caption{Distribution of stellar log probabilities for each cluster. Clusters shaded grey are sub-groups of the Taurus and Sco-Cen regions. Clusters are ordered from best fit and the top to worst fit at the bottom; some exhibit a clear bimodality.}
     \label{fig:cluster_violin}
\end{figure}

This test was designed to evaluate how well \texttt{ChronoFlow} fits the observational data for each cluster. We calculated the individual stellar log probabilities using the LOOCV models and have plotted these distributions for each cluster in Figure \ref{fig:cluster_violin}. The absolute $\mathcal{P}_{f}$ values for each cluster are not particularly meaningful; the probability densities are non-zero across our entire parameter space so we do not expect to achieve anything close to 1. But, the difference in $\mathcal{P}_{f}$ between clusters is meaningful. In particular, we note that four of the five clusters highlighted in \S\ref{subsec:res_cluster_ages} as anomalies (excluding M34) all have a relatively poor fit here as well. The bimodality present in some clusters (e.g., Hyades, NGC 2281, M37, and H Persei) is also meaningful, indicating that there is a better fitting group of stars and a poorer fitting group. In H Persei, this clearly corresponds to the group of slow rotators that fit the model well, and the over-density of fast rotators that is unexpected given its age. In M37 and NGC 2281, the poorer fitting groups seem to be overdensities of intermediate rotators at a wide range of colors lying in the gap between the slow and fast sequences. In Hyades, this appears to be due to intermediate rotators as well, but in this case they are restricted to a $(BP-RP)_0$ range of $\approx$ 2-3.

\subsection{Simulated data tests} \label{subsec:res_sim_test}

\begin{figure*}
    \centering
    \includegraphics[width=0.94\textwidth]{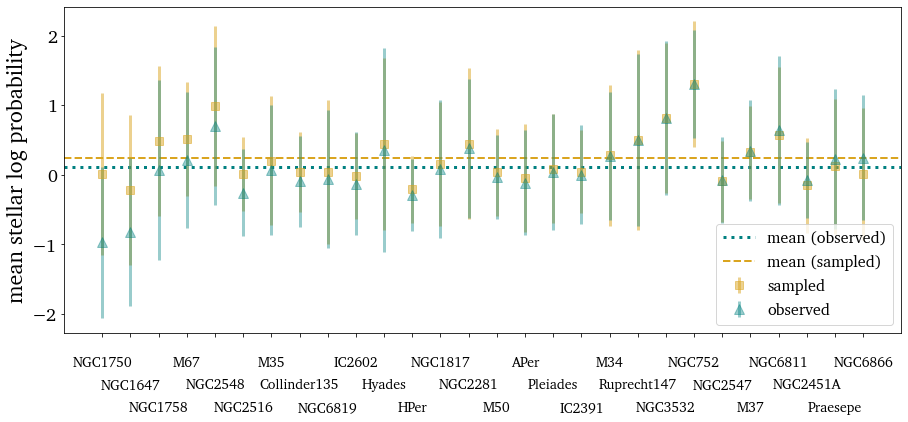}
    \caption{Comparison of the stellar log probability distributions for each cluster, calculated using observed data (teal triangles) and 200 samples (per observed star) simulated by \texttt{ChronoFlow} (gold squares). Clusters are ordered by difference between observed and sampled, where those with the poorest fit to observed data relative to sampled data are shown on the left.}
     \label{fig:benchmark_test}
\end{figure*}

A different test was required to evaluate \texttt{ChronoFlow}'s overall performance. To do this, we simulated data from the conditional probability densities that our model learned. If \texttt{ChronoFlow} could simulate data that closely \textbf{approximated} observations, we could be confident that it was learning the conditional probabilities well. In this test, we used our predicted probability density distributions from the LOOCV exercise to simulate rotation periods at the observed $(BP-RP)_0$, $p_{cl}$, $\tau$, and $\sigma_{C_0}$ values in our catalog. We drew 200 samples for each observed star, and compared the resulting log probabilities from our simulated data against those from the observations. Figure \ref{fig:benchmark_test} summarizes these results, averaging the log probabilities per cluster. Our model fits slightly better overall to the sampled data than the observations, however the difference is small relative to the scatter within each cluster. The difference is also cluster dependent; the five anomalous clusters discussed in \S\ref{subsec:res_cluster_ages} have larger differences on average than the others, but there are some clusters where the fit is actually better to the observations than the simulations. Given the scatter in this difference between clusters and between stars in a cluster, we conclude that \texttt{ChronoFlow} can model observed rotational evolution with high accuracy.

\begin{figure}[!htb]
    \centering
    \includegraphics[width=0.98\textwidth]{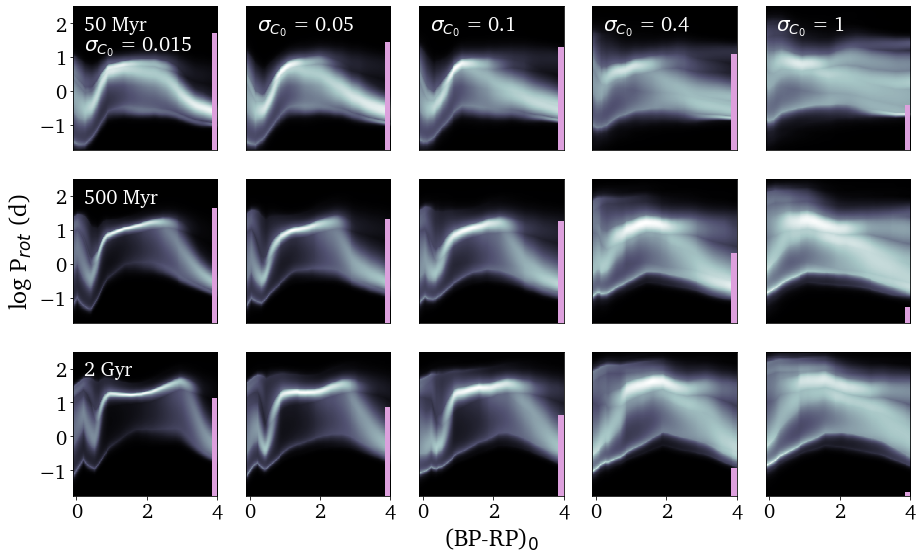}
    \caption{The effect of varying $\sigma_{C_0}$ on model predictions. Probability density slices in $(BP-RP)_0$ - $P_{rot}$ space are shown as a function of age and $\sigma_{C_0}$. The pink bars indicate the relative number of stars in each bin (scaled exponentially).}
     \label{fig:var_sigma0}
\end{figure}

\subsection{Impact of varying photometric uncertainty}
\label{subsec:res_var_sigma0}

Figure \ref{fig:var_sigma0} presents the results of varying $\sigma_{C_0}$ while keeping the other parameters constant for five different ages. As expected, greater photometric uncertainty smooths out the distribution. This is combination of two factors: (i) there is inherently more scatter in the $P_{rot}-C_0$ relationship when $\sigma_{C_0}$ is larger, and (ii) at large values of $\sigma_{C_0}$ there is less training data (see the pink bars), so the learned density distributions are naturally less constrained in that regime. Therefore, \texttt{ChronoFlow} will naturally have more uncertainty in age predictions when $\sigma_{C_0}$ is larger.

\section{Systematic Tests}
\label{app:systematic_details}

Here we provide additional details of the systematic tests and results that were summarized in \S\ref{sec: discussion}

\subsection{Dustmap comparison}
\label{subsec:sys_dustmap_comparison}

\begin{figure*}
    \centering
    \plotone{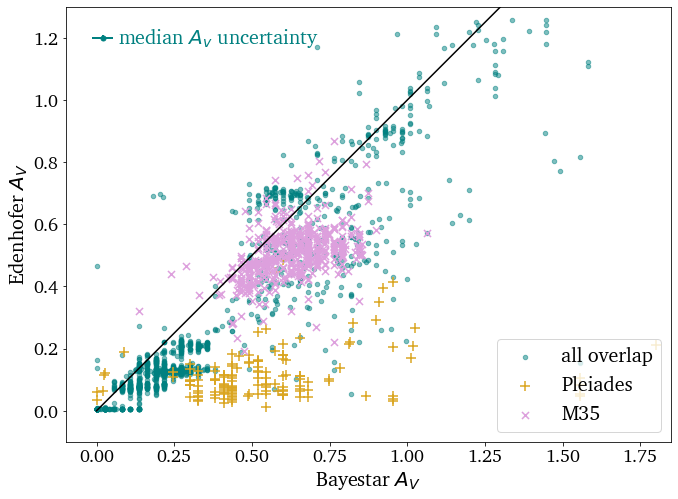}
    \caption{Comparison of the extinction on our catalog stars calculated using \cite{Edenhofer2023...dustmap...2023arXiv230801295E} and \cite{Bayestar19...dustmap...2019ApJ...887...93G}. The median uncertainty is shown in both directions in the upper left for reference (note that the Edenhofer uncertainty is much smaller than the Bayestar).}
    \label{fig:dustmap_overlap_comparison}
\end{figure*}

There is a systematic due to the choice of dustmap used to calculate extinctions. Extinction uncertainties are non-trivial due to: (i) parallax (and thus distance) uncertainties, (ii) sampling uncertainties within a single dustmap, and (iii) conversion uncertainties for each photometric band. They are the dominant source of photometric uncertainty in our final data catalog (compared to Gaia systematics and DR3 BP and RP uncertainties). In Figure \ref{fig:dustmap_overlap_comparison} we present a comparison of $A_v$ calculated from the \citet{Edenhofer2023...dustmap...2023arXiv230801295E} (E23) and \citet{Bayestar19...dustmap...2019ApJ...887...93G} (B19) dustmaps for stars in our catalog that are covered by both. The $A_V$ values obtained from the E23 map are systematically lower than those obtained from the B19 map. These results are qualitatively consistent with those found by \citet{GodoyRivera...KeplerDR3...2025arXiv250118719G}.
\begin{figure*}[t]
\centering
\begin{minipage}[t]{0.45\textwidth}
    \includegraphics[width=\textwidth]{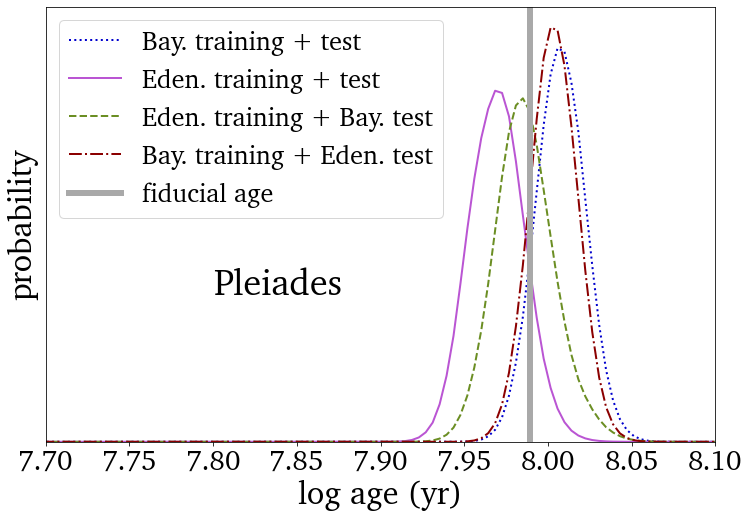}
    \caption{Cluster age posterior probabilities for the Pleiades for each of our four test cases comparing the E23 and B19 dustmaps.}
    \label{fig:Pleiades_dustmap_comparison}
\end{minipage}
\hfill
\begin{minipage}[t]{0.45\textwidth}
    \includegraphics[width=\textwidth]{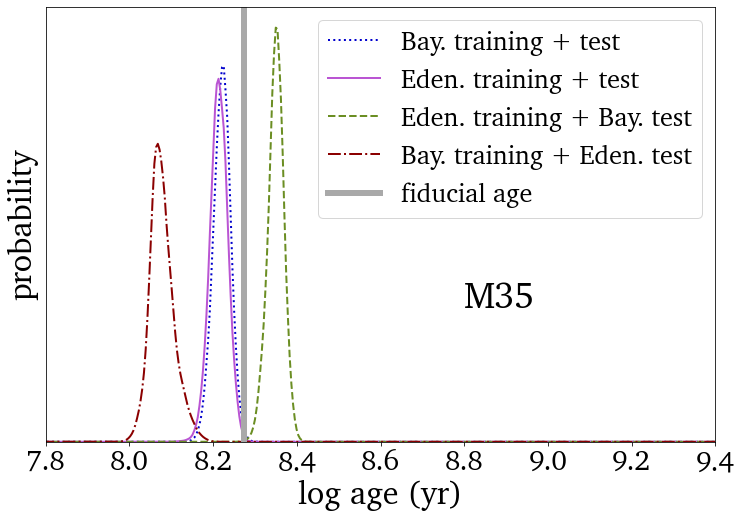}
    \caption{Cluster age posterior probabilities for M35 for each of our four test cases comparing the E23 and B19 dustmaps.}
    \label{fig:M35_dustmap_comparison}
\end{minipage}
\end{figure*}
To analyze the effect on our gyrochronology model, we examined two clusters in detail that have the largest differences, and which each have many stars covered by both maps: the Pleiades and M35. For each cluster, we generated a cluster posterior age probability distribution in four ways, essentially testing each combination of dustmaps for both training and testing:

\begin{enumerate}
    \item Using LOOCV, we trained a model using the E23 extinctions where there was overlap in coverage, and evaluated the cluster posteriors with E23 photometry.
    \item We trained a model using the E23 extinctions, but used B19 to infer the cluster ages.
    \item We trained a model using the B19 extinctions, but used E23 to infer the cluster ages.
    \item We trained a model using the B19 extinctions, and used B19 photometry to infer the cluster ages.
\end{enumerate}
The resulting posteriors for each test case are shown in figures \ref{fig:Pleiades_dustmap_comparison} and \ref{fig:M35_dustmap_comparison}. The scatter for all four tests for the Pleiades was small ($\sigma$ = 0.014 dex). The M35 age estimates are more scattered, however \textit{the age estimates were consistent when the training dust map was the same as the test dust map}. Analyzing the results in a different way, we calculated the scatter in age estimates for both clusters to be 0.01 dex when comparing test cases in which a single dustmap was used. The test cases where dustmaps were mixed resulted in an uncertainty of 0.04 dex. These indicate that the systematic uncertainty due to choice of dustmap is small, \textit{if the dustmap is used consistently}. However, caution should be taken when applying \texttt{ChronoFlow} to photometry de-reddened with a method other than the E23 or B19 maps.

\subsection{Source survey (space vs. ground)}
\label{subsec:disc_spaceonly}

We included observations from a variety of both space based and ground based source surveys in our data catalog. Overall, 36\% of our data are from ground surveys, and we do not have any space data for M34, M50, NGC 2281, NGC 2516, NGC 2548, NGC 3532, or NGC 752. Given the additional complications associated with observing from ground-based facilities (eg. atmospheric effects, discontinuous data due to the Earth's rotation), we wanted to determine whether ground data would introduce systematics not seen in only space data. To test this, we executed a LOOCV test using only rotation periods sourced from space surveys. The median cluster absolute residual increased by 3\%, and there was no apparent reduction in systematic uncertainties. This indicates that the value of adding more data from ground-based surveys outweighs the reduction in homogeneity of our catalog.

\begin{figure*}
    \centering
    \plotone{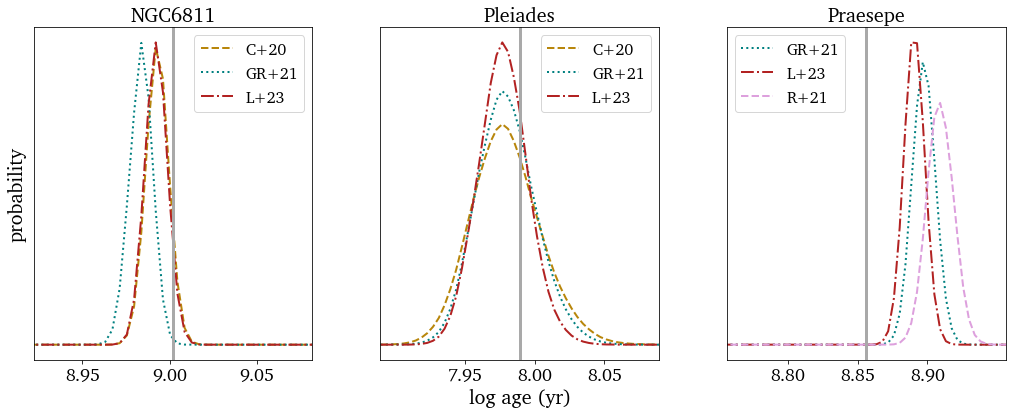}
    \caption{A comparison of the LOOCV results for three clusters when the test data were selected exclusively from individual membership catalogs: \citealt{Curtis2020_gyro} (C+20), \citealt{GodoyRivera...2021...rotcatalogue} (GR+21), \citealt{Long_gyrocatalogue...2023ApJS..268...30L} (L+23), and \citealt{Rampalli...Praesepe...2021ApJ...921..167R} (R+21). The vertical grey line represents the fiducial cluster age. Note that for NGC 6811, the C+20 and L+23 posteriors overlap almost completely. The scatter in inferred ages due to this effect is 0.01 dex.}
    \label{fig:dc_comparison}
\end{figure*}

\subsection{Cluster membership}
\label{subsec:disc_sys_dc_comparison}

Using our LOOCV models, we inferred ages using different observational data sets of the same cluster, from different literature cluster membership catalogs. Since we have three distinct literature catalogs for the Pleiades, Praesepe, and NGC 6811, we used these as test clusters.

Results are presented in Figure \ref{fig:dc_comparison}, which shows that there is little variance in the inferred ages ($\sigma$ = 0.01 dex). This suggests that overall, \texttt{ChronoFlow} is robust against cluster membership choices. However, it should be noted that these clusters are well sampled; different cluster membership datasets covering completely different parameter spaces may have more significant variation.

\subsection{Influence of training ages}
\label{subsec:disc_sys_training_ages}

\begin{figure*}
    \centering
    \plotone{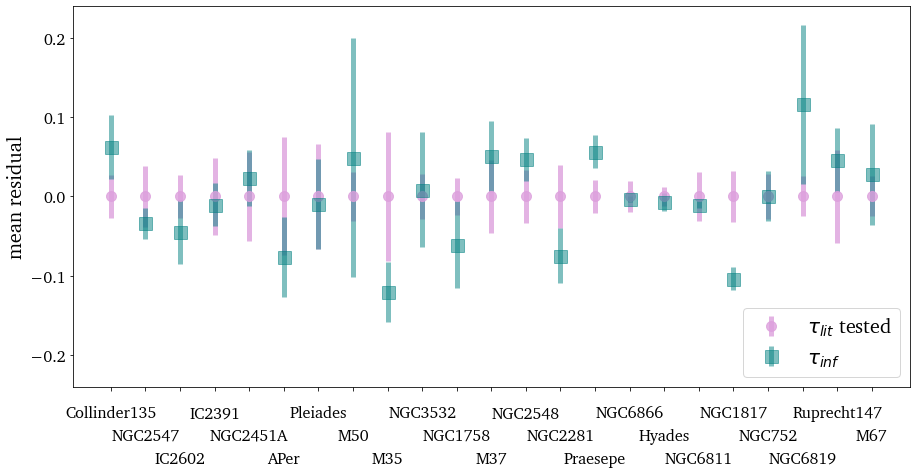}
    \caption{A comparison of the mean and variation ($1\sigma$ shown) in residuals from five different LOOCV models. The pink lines indicate the standard deviation in literature values used for each cluster across the models. The young sub-groups of Taurus and Scorpius are not shown here, nor are the five outlier clusters discussed in \S\ref{subsubsec:res_anomalous_clusters}.}.
    \label{fig:lit_age_test}
\end{figure*}

Since cluster ages are model dependent, it is important to examine the impact that varying calibration ages has. We did this by analyzing LOOCV results using five different version of \texttt{ChronoFlow}:

\begin{enumerate}
    \item The base model trained on all clusters excluding H Persei.
    \item We instead used Lithium depletion ages for the clusters that have such estimates available ($\alpha$ Persei, IC 2391, IC 2602, NGC 2451A, NGC 2547, and Pleiades). These estimates come from \cite{Stauffer...LiDep...Pleiades...1998ApJ...499L.199S}, \cite{Stauffer...LiDep...APer...1999ApJ...527..219S}, \cite{Oliveira...LiDep...NGC2547...2003MNRAS.342..651O}, \cite{JeffriesOliveira...LiDep...NGC2547...2005MNRAS.358...13J}, \cite{GutierrezAlbarran...LiDep...2020A&A...643A..71G}, \cite{GalindoGuil...LiDep...2022A&A...664A..70G}, and \cite{Jeffries...S.TESS.OC...2023MNRAS.526.1260J}. For clusters where multiple estimates were available, we took the geometric average.
    \item A model where we have taken the average age from \textit{all} literature sources described in table \ref{tab:all_lit_ages} instead of just the four primary catalogs.
    \item Ages directly from \cite{CantatGaudin...catalogue...2020AA...640A...1C}. Since they do not provide ages for Hyades or M34, we instead use the \cite{Bossini...Catalogue...2019AA...623A.108B} age for M34 and the \cite{GaiaCollab...2018AA...616A..10G} age for Hyades.
    \item A model where we prioritize the B+19 and G+18 ages where available, then L+23, then CG+20. We use the average of B+19 and G+18 for 11 clusters (where both have estimates), just B+19 for eight clusters, G+18 for four clusters, L+23 for two clusters, and CG+20 for the last three clusters.
\end{enumerate}

\noindent For the young sub-groups, we used the ages provided by \cite{Ratzenbock...ScoCen_groups...2023AA...677A..59R} and \cite{Krolikowski...Taurus...2021AJ....162..110K} for all models.

We performed this exercise to emulate the effect of model-dependent uncertainty in training ages (e.g., between lithium depletion, eclipsing binaries, isochrone families, asteroseismology). We compare the resulting variance in age inference relative to the variance in training ages in Figure \ref{fig:lit_age_test}. We also show the mean residual of the inferred age on the plot, but note that is a separated systematic captured by the LOOCV test, and here we are focused on the variance due to calibration ages. As expected, for age ranges where there is more variance in calibration ages, there is more variance in inferred residuals (e.g., at the ages of $\approx$ $\alpha$ Persei to NGC 3532). M50 and NGC 6819 are the most susceptible to the variation, with scatter exceeding 0.1 dex.

Overall, the age recovery variance is on the same order as the calibration variance. This indicates that varying training ages will have a similar order effect on the inferred ages in the model, which is $\approx$ 0.04 dex.

\end{document}